\documentclass[preprint,nofootinbib,tightenlines,superscriptaddress]{revtex4}
\usepackage{multirow,mathtools,amsmath,amssymb,graphicx,epsfig,amsfonts,color,epstopdf,adjustbox,lipsum,slashed,float,subfloat}

\definecolor{blu}{cmyk}{1,0.7,0,0.6}

\begin{document}

\baselineskip=15pt \parskip=3pt

\title{\color{blu} Right-handed neutrinos: Dark matter, lepton flavor violation, and leptonic collider searches}

\author{Meziane Chekkal}
\email{meziane.chekkal@univ-usto.dz, meziane.chekkal@gmail.com}
\affiliation{Department of Physics, University of Sciences and Technology of Oran,
B.P. 1505, Oran, El M'Naouer, Algeria}
\author{Amine Ahriche}
\email{aahriche@ictp.it, aahriche@gmail.com}
\affiliation{Department of Physics, University of Jijel, P.B. 98 Ouled Aissa, DZ-18000
Jijel, Algeria}
\affiliation{The Abdus Salam International Centre for Theoretical Physics, Strada
Costiera 11, I-34014, Trieste, Italy}
\author{Amine Bouziane Hammou}
\email{amine.hammou@univ-usto.dz, hbamine@gmail.com}
\affiliation{Department of Physics, University of Sciences and Technology of Oran,
B.P. 1505, Oran, El M'Naouer, Algeria}
\author{Salah Nasri}
\email{snasri@uaeu.ac.ae}
\affiliation{Department of Physics, United Arab Emirates University, P.O. Box 15551, Al-Ain, United Arab Emirates}

\begin{abstract}
We examine lepton flavor violating (LFV) interactions for heavy
right-handed neutrinos that exist in most of the standard
model extensions that address dark matter (DM) and neutrino mass
at the loop level. In order to probe the collider effect of these LFV
interactions, we impose the assumption that the model parameters give
the right values of the DM relic density and fulfill the constraints from the LFV processes 
$\ell_{\alpha}\rightarrow\ell_{\beta}\gamma$ and $\ell_{\alpha}\rightarrow3\ell_{\beta}$. 
We also investigate the possibility of probing these interactions, and hence the right-handed neutrino, 
at leptonic colliders through different final state signatures.
\\
\\

\end{abstract}
\maketitle

\section{Introduction}

The standard model (SM) of particle physics describes the properties of
elementary particles and their interactions and has been in good agreement
with numerous experiments, so far. Moreover, one of its last building
blocks, namely, the Higgs boson, has been detected, and there are future
planed experiments to precisely measure its couplings to matter as
well as its self-coupling. Neutrino oscillation data established that
at least two of the three SM neutrinos have mass
and nonzero mixing. However, their properties, such as their nature
and the origin of the smallness of their mass, have no explanation
within the SM, which cry out for new physics. One of the most popular
mechanisms for understanding why neutrino mass is tiny is the seesaw
mechanism~\cite{seesaw}, which assumes the existence of right-handed
(RH) neutrinos with masses several orders of magnitude heavier than 
the electroweak (EW) scale. In this approach, the dimension five operator
$(L\Phi)^{2}/\Lambda$ will be induced in the low-energy effective
Lagrangian, with $\Lambda$ the scale of the new physics which
is of the order of the RH neutrino mass scale. Unfortunately, such heavy
particles decouple from the low-energy spectrum at energies many
orders of magnitude higher than the EW scale and so cannot be
directly probed in high-energy physics experiments. Even in a low-scale seesaw mechanism where 
$\Lambda \sim ~\text{TeV}$, one typically expects the active-singlet neutrino mixing to be smaller than $10^{-5}$, 
making the production cross section well below the current and the near future collider 
sensitivities~\cite{Lowseesaw}. Moreover, for a RH neutrino
heavier than $10^{7}~\text{GeV}$, it can destabilize the electroweak
vacuum via the large loop corrections induced by the Dirac neutrino
Yukawa term. 

Another possible way to explain the smallness of neutrino masses
is by generating them at the loop level~\cite{Zee,ZB,Ma,KNT,Aoki:2008av}. In
this mechanism, the tree-level neutrino mass term vanishes due to some
discrete symmetry and can be induced via loop diagrams with a loop
suppression factor, and, consequently, the scale of new physics $\Lambda$
can be much lower than the conventional seesaw mechanism. For instance,
as it has been shown in Ref.~\cite{AMN}, the scale of new physics can
be in the sub-TeV for the three-loop neutrino mass generation model~\cite{KNT},
which makes it testable at collider experiments~\cite{ANS}.

In most of the radiative neutrino mass models, neutrinos are Majorana
particles, and so one expects a violation of the total lepton number.
This will have stringent phenomenological implications on loop-induced
neutrino mass models and can be a direct indication of new physics.
Indeed, lepton flavor violating (LFV) processes are currently the
object of great attention; the experiments which aim at detecting
them are becoming increasingly precise and clean. The radiative decay 
$\mu\rightarrow e\gamma$ is simplest to detect; indeed, only one particle is created in the finale state, the electron.
Its energy is thus of the order of $m_{\mu}=105~\text{MeV}$, which is far from the principal background
$\mu\rightarrow e\bar{\nu_{e}}\nu_{\mu}$ disintegration, whose energy spectrum decreases
drastically beyond $m_{\mu}/2$. The current experimental limits for
these low-energy processes are $\mathcal{B}\left(\mu\rightarrow e\gamma\right)<5.7\times10^{-13}$
(MEG~\cite{mueg}), $\mathcal{B}\left(\mu\rightarrow e^{-}e^{+}e^{-}\right)<10^{-12}$
(SINDRUM~\cite{mue}), $\mathcal{B}\left(\tau\rightarrow e\gamma\right)<3.3\times10^{-8}$
(\textit{BABAR}~\cite{taug}), $\mathcal{B}\left(\tau\rightarrow e^{-}e^{+}e^{-}\right)<2.7\times10^{-8}$
(BELLE~\cite{tau}), $\mathcal{B}\left(\tau\rightarrow\mu\gamma\right)<4.4\times10^{-8}$
(\textit{BABAR}~\cite{taug}) and $\mathcal{B}\left(\tau\rightarrow\mu^{-}\mu^{+}\mu^{-}\right)<2.1\times10^{-8}$
(BELLE~\cite{tau}), and some of these bounds are expected to improve
by almost an order of magnitude in the next couple of years. In particular,
the MEG experiment will be sensitive to a $\mu\rightarrow e\gamma$
branching ratio as low as $6\times10^{-14}$ which will be able
to strongly constrain a large class of radiative neutrino mass models.

In this work, we study the phenomenological implications of a class
of radiative neutrino mass models based on the SM extension by RH neutrinos and a singlet charged scalar and in which the lightest
RH neutrino is a dark matter candidate. After imposing the current
bounds from the LFV processes on the model parameters, we determine the parameters space
for which the RH relic density is in agreement with the observation.
We also consider the constraint on single- and multiphoton events
with missing energy reported by the L3 Collaboration at LEP~\cite{LEP}. In our scan, we find that many 
of the benchmark points that satisfy the LFV constraints require a cancellation among the product 
combinations of the coupling to cancel out. For that, we define a quantify that quantifies such a 
fine-tuning parameter and use it to classify the viable model parameter space. Finally, we study 
the possibility of observing a signal at future lepton colliders.

The paper is organized as follows: In Sec. II, we present
a class of interactions involving right-handed neutrinos, that appear in many radiative neutrino 
mass models, and discuss the constraints from the different LFV processes. In Sec. III, we show 
that the lightest RH neutrino can be a viable dark matter candidate while satisfying the current 
experimental bounds on LFV processes. In Sec. IV, we analyze the monophoton events with missing 
energy using the data collected by the L3 Collaboration at LEP-II and determine the viable parameter 
space for such models. In Sec. V, we consider three benchmark points (according to the fine-tuning 
degree) and study the possible signature of the RH neutrino and charged scalar signals at lepton colliders. 
We also discuss the effect of using polarized beams on the signal significance. Finally, we give our conclusion in Sec. VII.

\section{LFV Constrains Class of Models with right-handed Neutrinos}

\begin{figure}[htp]
\begin{centering}
\includegraphics[width=14cm,height=4cm]{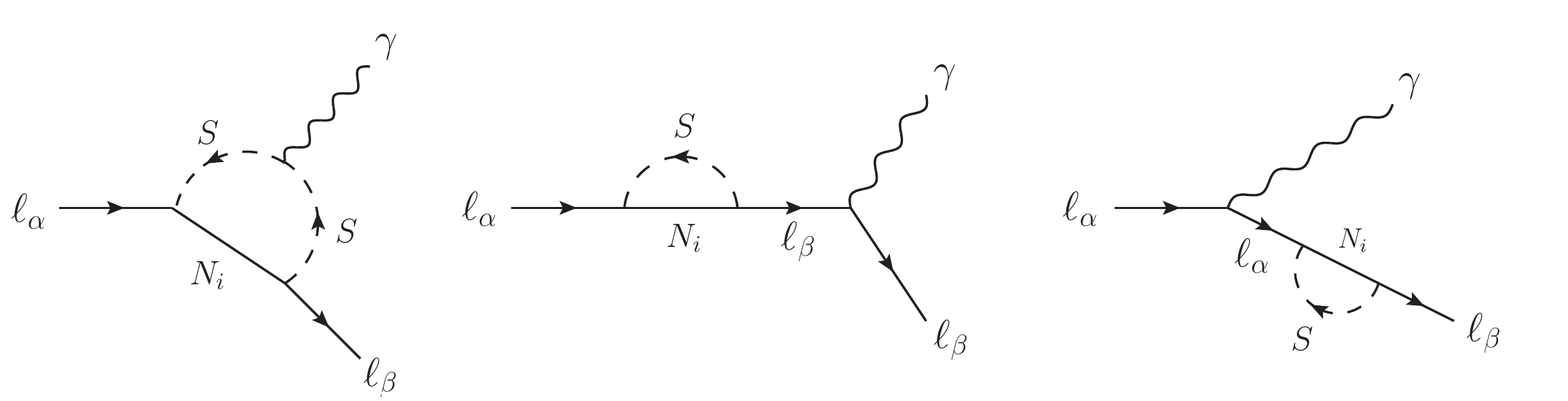} 
\par\end{centering}
\caption{The one-loop diagrams that contribute to $\ell_{\alpha}\rightarrow\ell_{\beta}\gamma$.}
\label{one}
\end{figure}

\begin{figure}[htp]
\begin{centering}
\includegraphics[width=14cm,height=6.5cm]{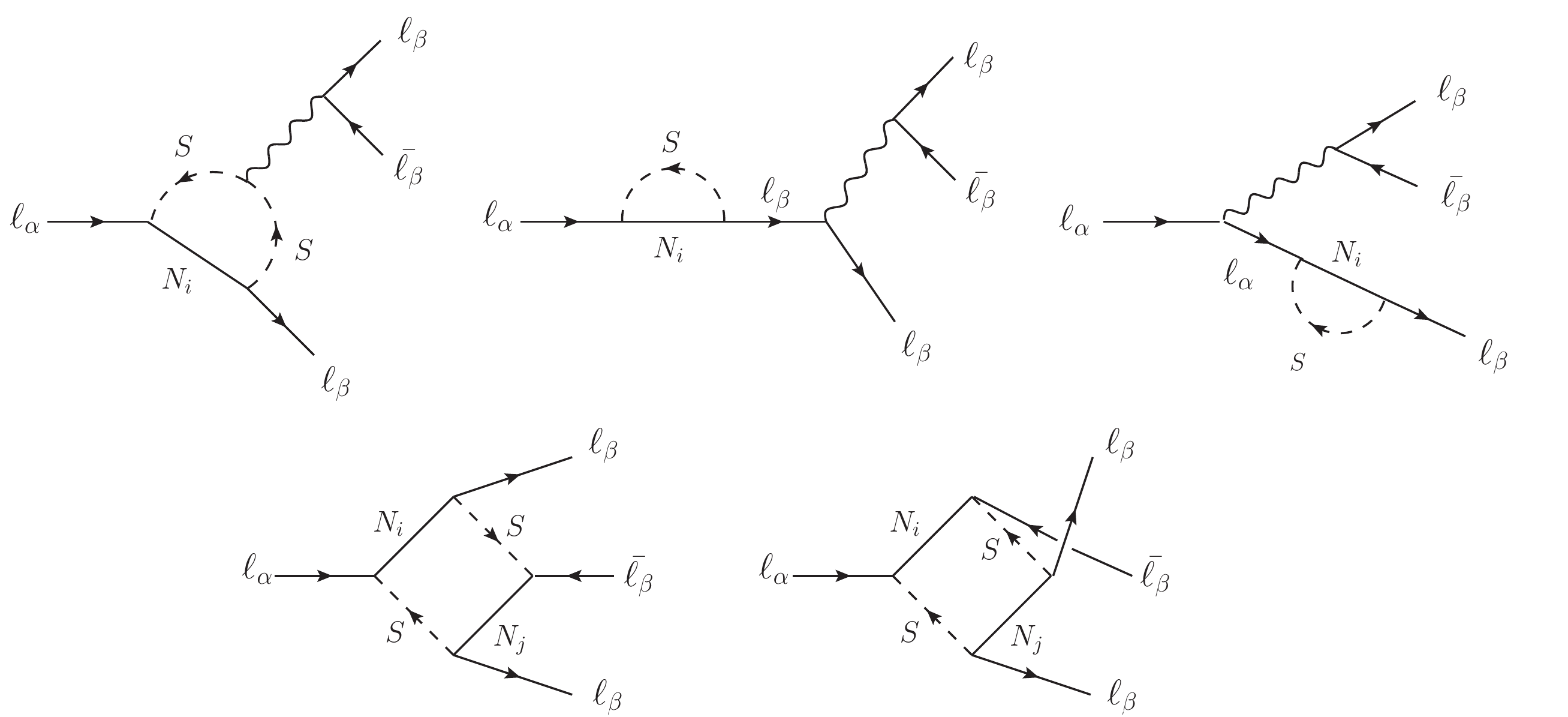} 
\par\end{centering}
\caption{The diagrams contributing to $\ell_{\alpha}\rightarrow 3\ell_{\beta}$.
The penguin (box) diagrams are shown in the top (bottom).}
\label{three} 
\end{figure}

We consider a class of radiative neutrino mass models based on extending
the SM with three right-handed neutrinos $N_{i}~(i=1,2,3)$
and a charged scalar field $S^{\pm}$ which is an $SU(2)_{L}$ singlet.
For the purpose of our study, the relevant term in the Lagrangian
has the form~\cite{Ma, KNT, Aoki:2008av, Okada:2015hia}
\begin{equation}
\mathcal{L_{N}}\supset-\frac{1}{2}m_{N_{i}}\overline{N_{i}^{c}}P_{R}N_{i}+g_{i\alpha}S^{+}\overline{N_{i}}\ell_{\alpha_{R}}+\mathrm{H.c.},\label{LL}
\end{equation}
where $\ell_{\alpha_{R}}$ is the right-handed charged lepton and $g_{i\alpha}$ are Yukawa couplings. The global
$Z_{2}$ symmetry is imposed\footnote{In some neutrino mass models like Ref.~\cite{Toma}, where the RH neutrino are 
promoted to higher representations (septplet), the global $Z_2$ symmetry is accidental.} to ensure the stability 
of the lightest RH neutrino, that is supposed to play the dark matter (DM) role. These interactions can give rise to the LFV processes $\ell_{\alpha}\rightarrow \ell_{\beta}\gamma(\{\alpha=\mu,\beta=e\})$ and $\ell_{\alpha}\rightarrow 3\ell_{\beta}(\{\alpha=\tau,\beta=e,\mu\})$,
that are mediated by the RH neutrinos and the charged scalar fields as shown
in Figs.~\ref{one} and ~\ref{three}, respectively.

The contribution of the interactions (\ref{LL}) to the branching ratio $\ell_{\alpha}\rightarrow\ell_{\beta}\gamma$ is~\cite{Toma:2013zsa}

\begin{equation}
\mathcal{B}^{(N)}(\ell_{\alpha}\rightarrow\ell_{\beta}\gamma)=\frac{3(4\pi)^{3}\alpha}{4G_{F}^{2}}|A_{D}|^{2}\times\mathcal{B}(\ell_{\alpha}\rightarrow\ell_{\beta}\nu_{\alpha}\bar{\nu}_{\beta}),\label{meg}
\end{equation}
where $G_{F}$ is the Fermi constant, $\alpha=e^{2}/4\pi$ is the
electromagnetic fine structure constant, and $A_{D}$ is the dipole
contribution that is given by

\begin{equation}
A_{D}=\sum_{i=1}^{3}\frac{g_{i\beta}^{\ast}g_{i\alpha}}{2(4\pi)^{2}}\frac{1}{m_{S}^{2}}F\left(x_{i}\right),\label{AD}
\end{equation}
with $x_{i}=m_{N_{i}}^{2}/m_{S}^{2}$ and $F(x)$ is a loop function given in the Appendix.

For $\ell_{\alpha}=\ell_{\beta}=\mu$, Fig.~\ref{one} represents a new contribution to the muon anomalous magnetic moment $\delta a_{\mu}$, and it is given by

\begin{align}
\delta~a_{\mu}^{(N)} =\frac{1}{16 \pi^2}\frac{m_{\mu}^2}{m_S^2} \sum_i \left|g_{i\mu}\right|^2F_2\left(\frac{m_{N_i}^2}{m_S^2}\right).\label{damu}
\end{align}

The branching ratio for $\ell_{\alpha}\rightarrow\ell_{\beta}\bar{\ell}_{\beta}\ell_{\beta}$
is~\cite{Toma:2013zsa} 

\begin{align}
\mathcal{B}^{(N)}(\ell_{\alpha}\rightarrow\ell_{\beta}\bar{\ell}_{\beta}\ell_{\beta}) & = \frac{3(4\pi)^{2}\alpha^{2}}{8G_{F}^{2}}\Bigg[|A_{ND}|^{2}+|A_{D}|^{2}\left(\frac{16}{3}\log\left(\frac{m_{\alpha}}{m_{\beta}}\right)-\frac{22}{3}\right)\nonumber\\
& +\frac{1}{6}|B|^2 +\frac{1}{3}\frac{m_{\alpha}^{2}m_{\beta}^{2}\left(3\sin^{4}\theta_{W}-\sin^{2}\theta_{W}+\frac{1}{4}\right)}{m_{W}^{4}\sin^{4}\theta_{W}}\left|A_{D}\right|^{2}\nonumber\\
& +\left(-2A_{ND}A_{D}^{\ast}+\frac{1}{3}A_{ND}B^{\ast}-\frac{2}{3}A_{D}B^{\ast}+\mathrm{h.c.}\right)\Bigg] \nonumber\\
& \times\mathcal{B}(\ell_{\alpha}\rightarrow\ell_{\beta}\nu_{\alpha}\bar{\nu}_{\beta}).\label{tmug}
\end{align}

Here $\theta_{W}$ is the Weinberg mixing angle. The coefficients $A_{ND}$ and $B$ are the nondipole contribution
from the photonic penguin and the box diagrams, respectively, which read

\begin{equation}
A_{ND}=\sum_{i=1}^{3}\frac{g_{i\beta}^{\ast}g_{i\alpha}}{6(4\pi)^{2}}\frac{1}{m_{S}^{2}}G\left(x_{i}\right),\label{AND}
\end{equation}
and
\begin{equation}
B=\frac{1}{(4\pi)^{2}e^{2}m_{S}^{2}}\sum_{i,j=1}^{3}\left[\frac{1}{2}D_{1}\left(x_{i},x_{j}\right)g_{j\beta}^{\ast}g_{j\beta}g_{i\beta}^{\ast}g_{i\alpha}+\sqrt{x_{i}x_{j}}D_{2}\left(x_{i},x_{j}\right)g_{j\beta}^{\ast}g_{j\beta}^{\ast}g_{i\beta}g_{i\alpha}\right],\label{B}
\end{equation}
where $G(x)$, $D_{1}\left(x_{i},x_{j}\right)$, and $D_{2}\left(x_{i},x_{j}\right)$
are loop functions given in the Appendix.

The contribution of the photonic dipole term is present in both branching
ratios and is more important than the photonic nondipole term regardless
of the values of the couplings and the RH neutrinos and the charged scalar masses. 
It is worth pointing out that it is possible to chose the model parameters for 
which the branching ratio of the trilepton channel is larger than the one for the $\ell_{\alpha}\rightarrow\ell_{\beta}\gamma$
channel where the main contribution comes from the box diagrams in Fig.~\ref{three}.

We perform a numerical scan over all free parameters of the model to probe 
possible signatures of new physics at colliders. In addition to the LFV constraints, we 
require the Yukawa couplings $g_{i\alpha}$ to be perturbative. In order to 
avoid the bounds on $\ell_{\alpha}\rightarrow \ell_{\beta}+\gamma$, one has to consider a small value for 
$g_{i\alpha}$ or have a cancellation between the different terms in the expression of $A_D$ (\ref{AD}). 
To quantify the fine-tuning that ensures such cancellation, we define
\begin{equation}
\mathrm{\textit{R}}=\frac{\mid\sum_{i=1}^{3}g_{i\beta}^{\ast}g_{i\alpha}F\left(x_{i}\right)\mid^{2}}{Max[\mid g_{i\beta}^{\ast}g_{i\alpha}F\left(x_{i}\right)\mid^{2}]},\label{R}
\end{equation}
which we will call the fine-tuning parameter. In this case, very small \textit{R} corresponds to a severe tuning 
on the model parameters. This could allow the Yukawa coupling $g_{i\alpha}$ to be large, which is interesting for 
collider searches. In what follows, we consider the fine-tuning 
parameter only for the process $\mu \rightarrow e+\gamma$ (i.e., $\alpha=\mu$ and $\beta=e$), since this is the most 
severely constrained.

\begin{figure}[htp]
\begin{centering}
\includegraphics[width=0.33\textwidth]{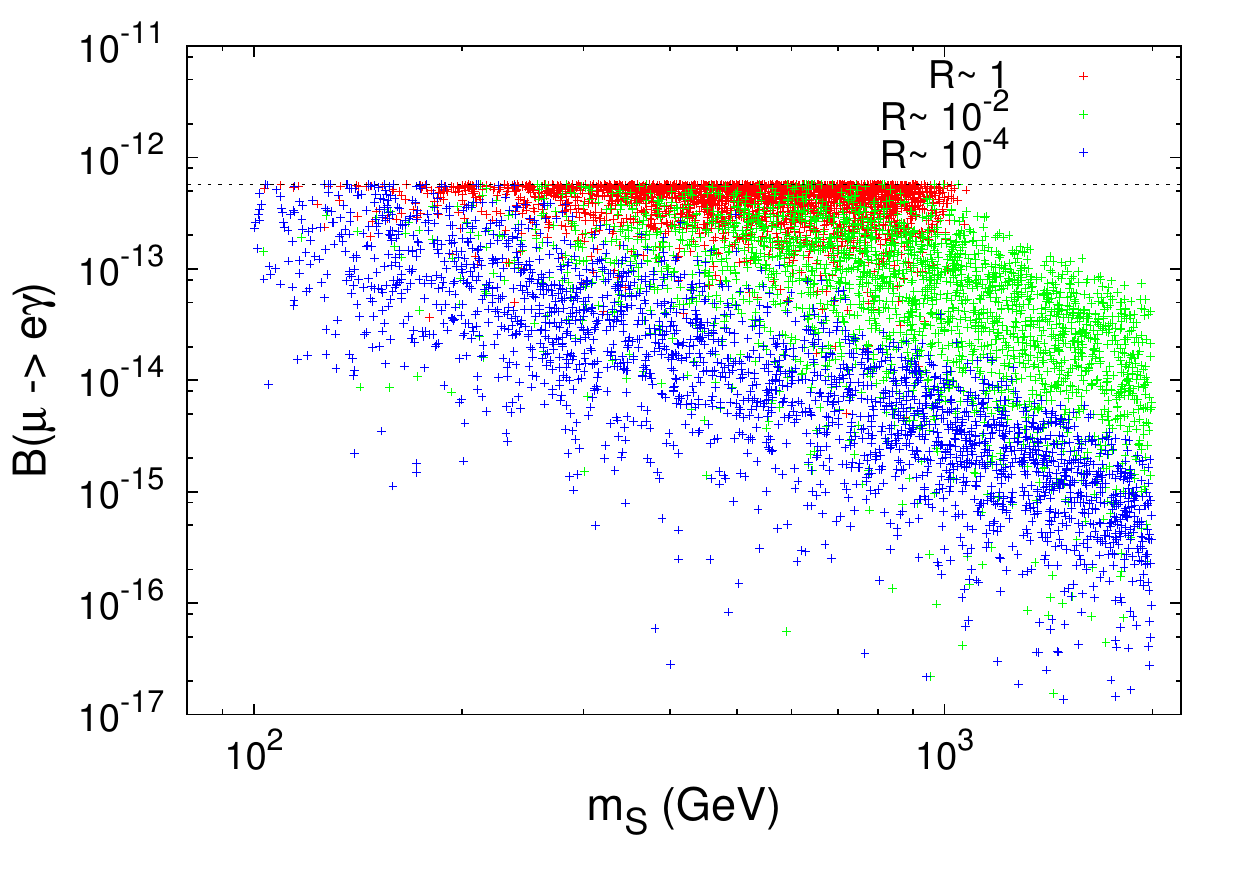}~\includegraphics[width=0.33\textwidth]{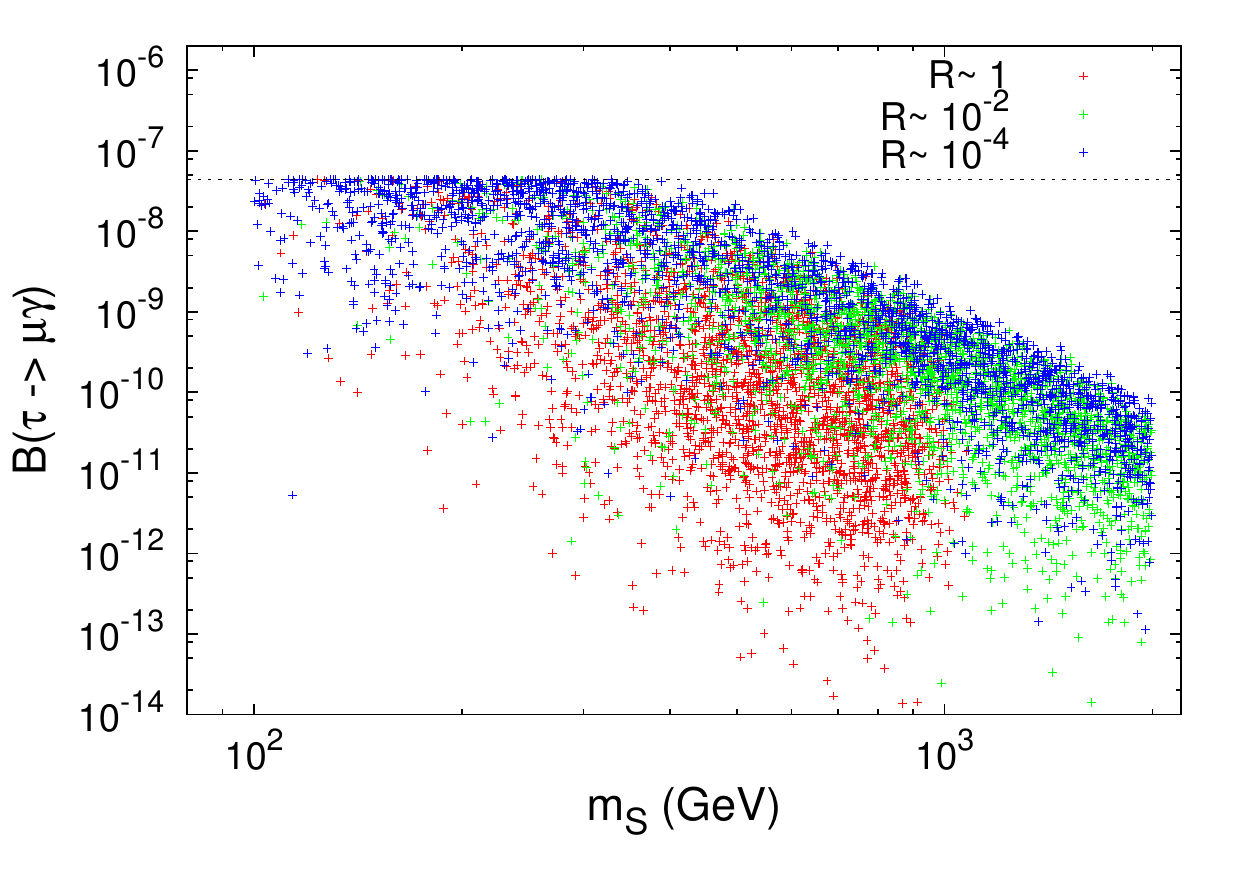}~\includegraphics[width=0.33\textwidth]{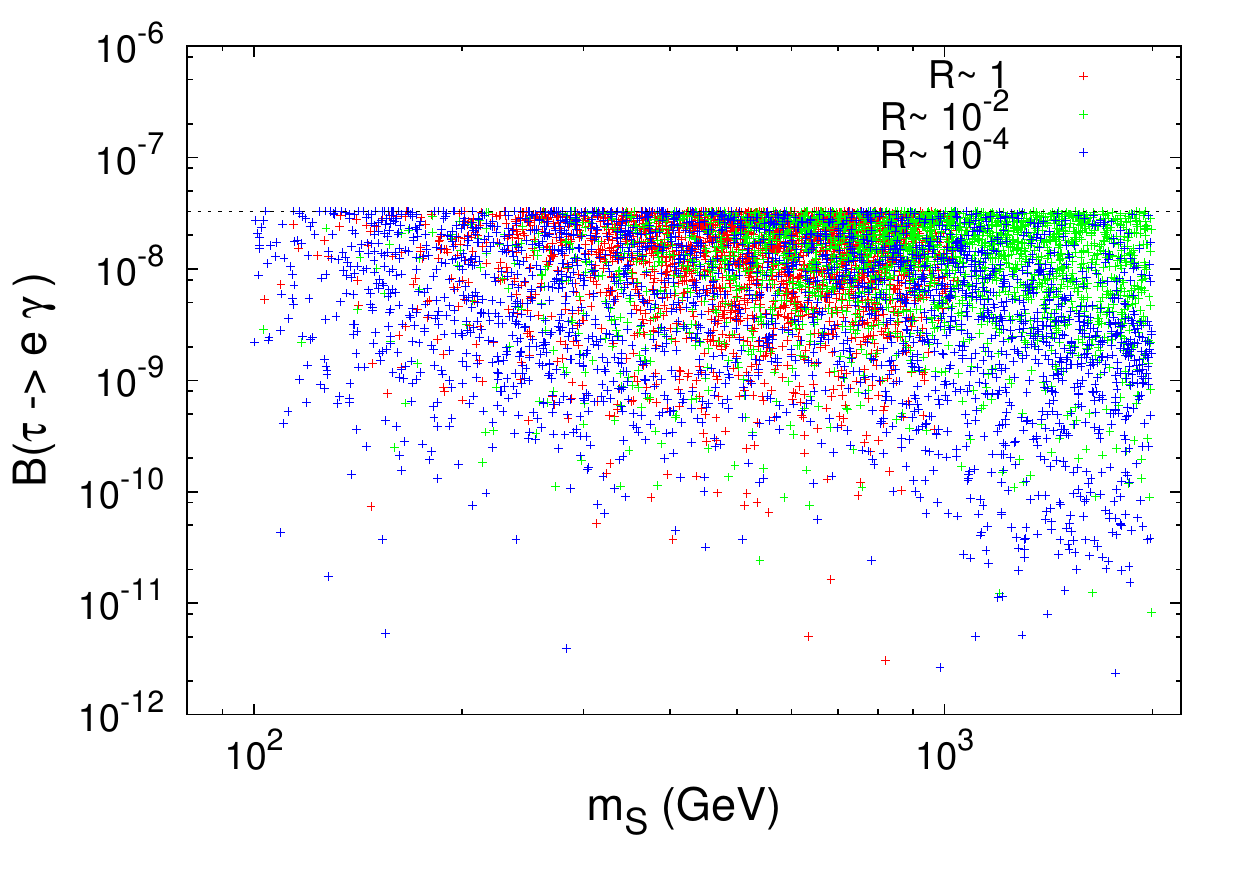}\\
\includegraphics[width=0.33\textwidth]{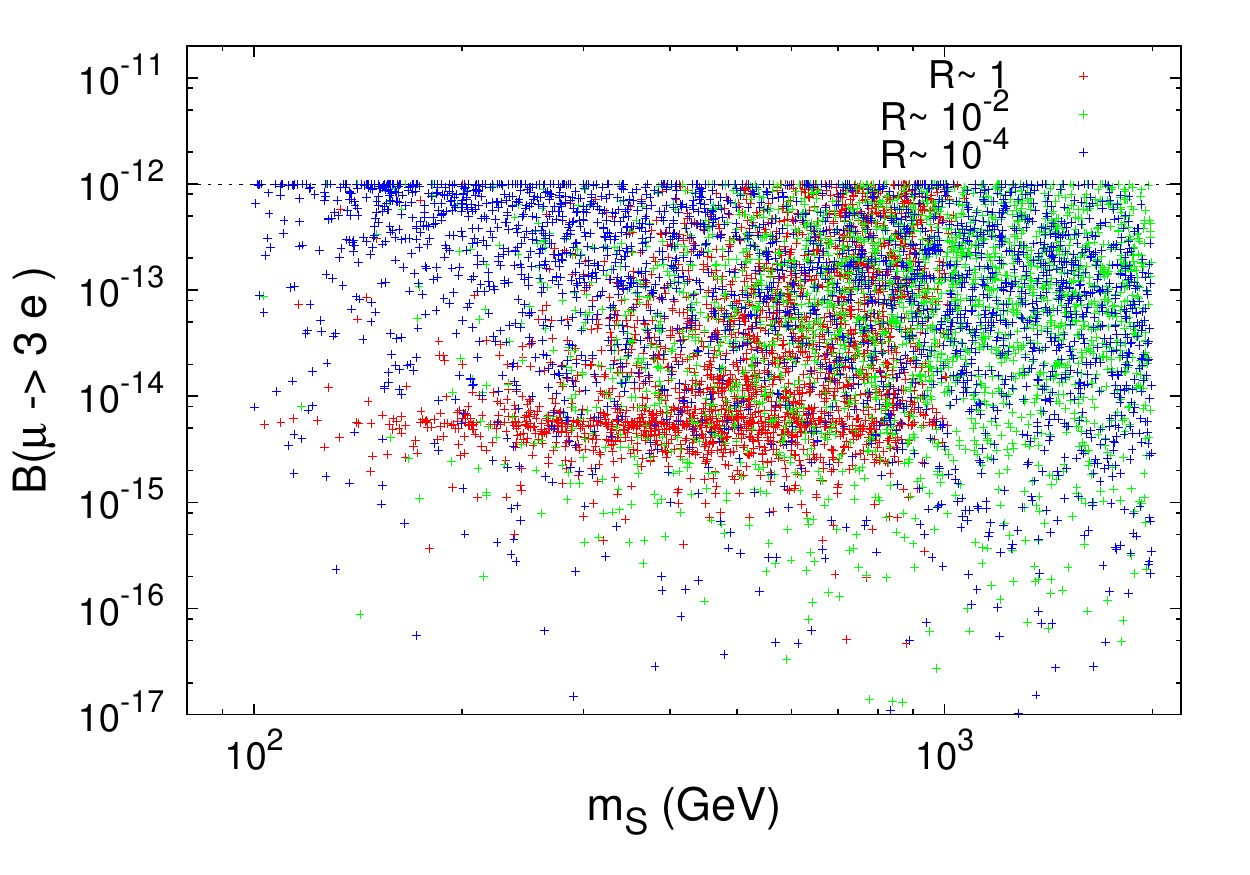}~\includegraphics[width=0.33\textwidth]{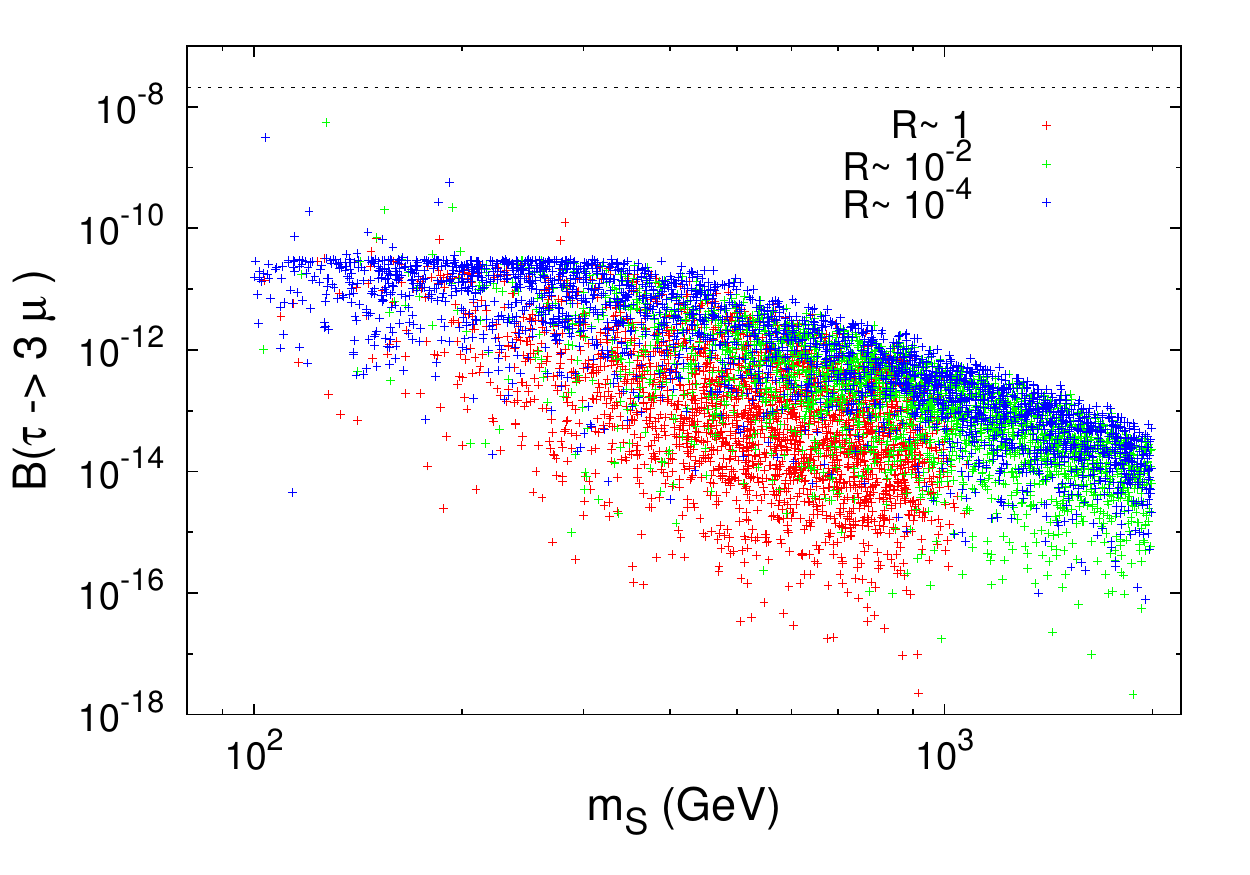}~\includegraphics[width=0.33\textwidth]{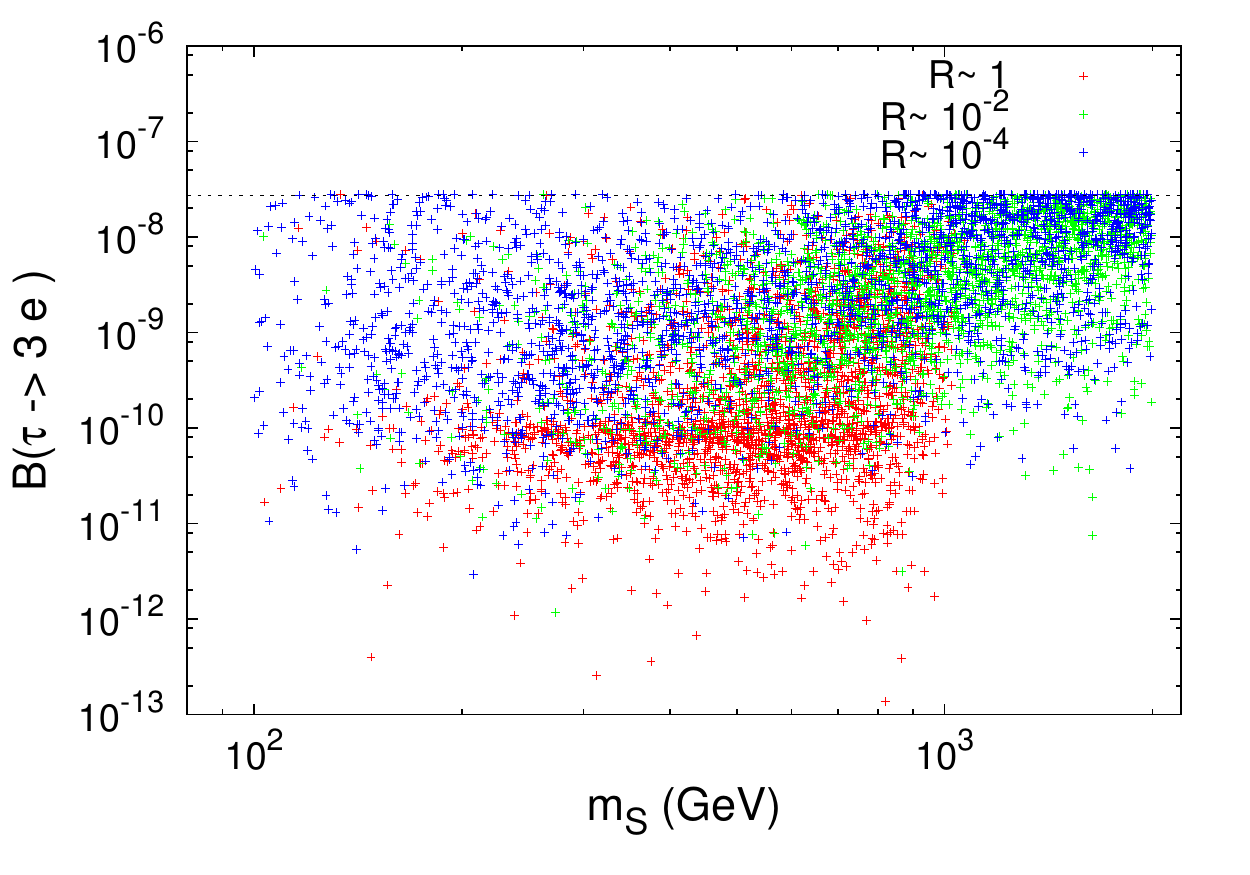}
~\includegraphics[width=0.33\textwidth]{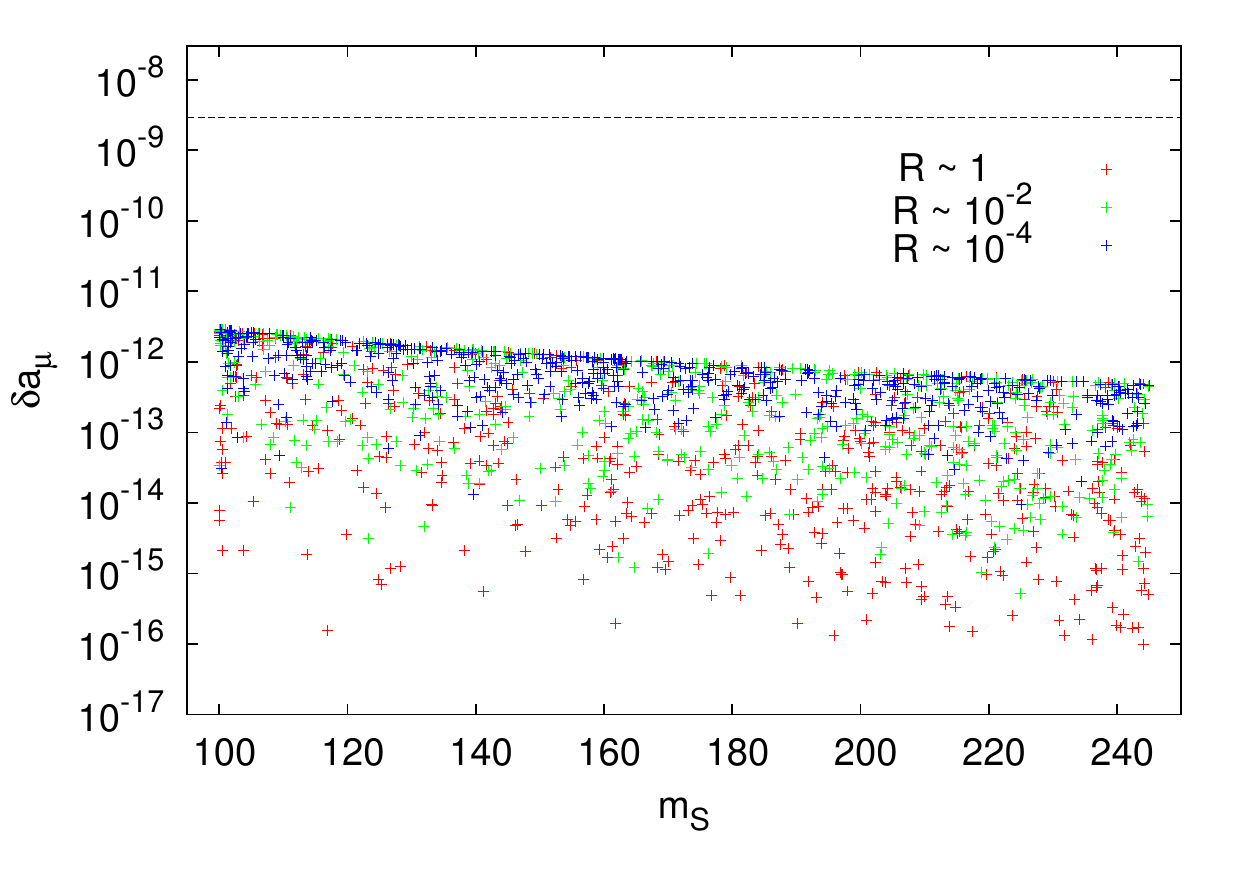}
\par\end{centering}
\caption{The branching ratios (top) $\mathcal{B}(\mu\rightarrow e\gamma)$, $\mathcal{B}(\tau\rightarrow\mu\gamma)$ and $\mathcal{B}(\tau\rightarrow e\gamma)$; and (middle) $\mathcal{B}(\mu\rightarrow 3e)$, $\mathcal{B}(\tau\rightarrow 3\mu)$, and $\mathcal{B}(\tau\rightarrow 3e)$ as a function of $m_{S}$. Some hundreds of configurations of free parameters
satisfying the experimental data are used. In the bottom, we show the contribution to the muon anomalous magnetic moment. The horizontal 
dashed lines show the current experimental upper bounds for these observables.}
\label{gg} 
\end{figure}

In Fig.~\ref{gg}, we present the branching ratios for the processes $\ell_{\alpha}\rightarrow\ell_{\beta}\gamma$, 
$\ell_{\alpha}\rightarrow 3 \ell_{\beta}$, and the anomalous magnetic moment of the muon as a function of the charged 
scalar mass for different values of the ratio $R\approx ~1,~10^{-2},~10^{-4}$. We note that although the most stringent 
constraint on LFV from the process $\mu \rightarrow e\gamma$ can be fulfilled for $R_1\approx 1$ by taking the 
couplings $g_{ie}$ and $g_{i\mu}$ small, this may be in conflict with the DM relic density for some values of 
the charged scalar masses, whereas for $R_2\approx ~10^{-2}$ and $R_3\approx ~10^{-4}$ the choice of ratio has a minor 
impact on the LFV branching fractions of the muon and tau decay processes.

As can be seen, for scalar and the RH neutrinos lighter than few hundreds GeV, the LFV decay processes can be in agreement with the experimental bounds. For the muon anomalous magnetic moments, the Yukawa interaction terms $S\bar{N}\ell_R$ give a 
contribution smaller than $10^{-12}$, and, hence, does not account for the $3.6\sigma$ deviation from the SM prediction~\cite{deltaa}, 
and require new physics if the discrepancy will be confirmed at the $5\sigma$ level by the upcoming experiments~\cite{Futureamu}.

In this class of models, the interactions term (\ref{LL}) does not induce a contribution to neutrinoless double beta decay, 
since both $N_R$ and $S^{\pm}$ do not couple to the quarks. However, if these interactions are embedded within a larger gauge 
symmetry, such as left-right symmetric models $\left[SU(2)_L \times SU(2)_R\right]$~\cite{Tello:2010am}, then there will be a contribution 
which involves the $W_R^{\pm}$ and depends on $M_{W_R}$, $m_N$, and the new gauge coupling strength. Depending on the model details, when 
light Majorana neutrino masses $m^{(\nu)}_{\alpha\beta}$ are generated, there will be contributions proportional to the 
element $m^{(\nu)}_{ee}$, for which the bound on the rate of $0\nu\beta\beta$ can be easily satisfied.

\section{Dark Matter: Relic Density}

As mentioned earlier, the lightest RH neutrino $N_{1}$ is stable
and could be a good DM candidate. In the hierarchical RH neutrino
mass spectrum case, we can safely neglect the effect of $N_{2}$ and
$N_{3}$ for $N_{1}$ density. The $N_{1}$ number density gets depleted
through the annihilation process $N_{1}N_{1}\rightarrow\ell_{\alpha}\ell_{\beta}$
via \textit{t}-channel exchange of $S^{\pm}$. For two incoming DM
particles with momenta $p_{1}$ and $p_{2}$, and final state charged
leptons with momenta $k_{1}$ and $k_{2}$, the amplitude is given
by

\begin{equation}
\mathcal{M}_{\alpha\beta}=g_{1\alpha}g_{1\beta}^{*}\left[\frac{\overline{u}\left(k_{1}\right)P_{L}u\left(p_{1}\right).\overline{v}\left(p_{2}\right)P_{R}v\left(k_{2}\right)}{t-m_S^{2}}-\frac{\overline{u}\left(k_{1}\right)P_{L}u\left(p_{1}\right).\overline{v}\left(p_{2}\right)P_{R}v\left(k_{2}\right)}{u-m_S^{2}}\right],\label{rd1}
\end{equation}
where $t=\left(p_{1}-k_{1}\right)^{2}$ and $u=\left(p_{1}-k_{1}\right)^{2}$
are the Mandelstam variables corresponding to the $t$ and $u$ channels, respectively.
After squaring, summing, and averaging over the spin states, we find
that, in the nonrelativistic limit, the total annihilation cross section
reads\footnote{In some models like Ref.~\cite{Ahriche:2016cio}, there are some annihilation channels beside $N_{1}N_{1}\rightarrow\ell_{\alpha}\ell_{\beta}$. In this case, the annihilation cross section of $N_{1}N_{1}\rightarrow\ell_{\alpha}\ell_{\beta}$ should be smaller than (\ref{rd2}), and the combination $\sum_{\alpha,\beta}|g_{1,\alpha}g_{1,\beta}^{*}|^{2}$ should be less than its value in (\ref{rd}).}
\begin{equation}
\sigma_{N_{1}N_{1}}v_{r}\simeq\sum_{\alpha,\beta}|g_{1,\alpha}g_{1,\beta}^{*}|^{2}\frac{m_{N_{1}}^{2}\left(m_{S}^{4}+m_{N_{1}}^{4}\right)}{48\pi\left(m_{S}^{2}+m_{N_{1}}^{2}\right)^{4}}v_{r}^2,\label{rd2}
\end{equation}
with $v_{r}$ the relative velocity between the annihilating $N_{1}$
particles. As the temperature of the Universe drops below the freeze-out
temperature $x_f=m_{N_{1}}/T_{f}\approx ~25$, the annihilation rate becomes
smaller than the expansion rate (the Hubble parameter) of the Universe,
and the $N_{1}$'s start to decouple from the thermal bath. The relic
density after the decoupling can be obtained by solving the Boltzmann
equation, which approximately yields~\cite{Ahriche:2013zwa}
\begin{equation}
\Omega_{N_{1}}h^{2}\simeq\frac{2x_{f}\times 1.1\times10^{9}~\text{GeV}^{-1}}{\sqrt{g^{*}}M_{pl}\left\langle \sigma_{N_{1}N_{1}}v_{r}\right\rangle }\simeq\frac{17.56}{\sum_{\alpha,\beta}|g_{1\alpha}g_{1\beta}^{*}|^{2}}\left(\frac{m_{N_{1}}}{50~\text{GeV}}\right)^{2}\frac{\left(1+m_{S}^{2}/m_{N_{1}}^{2}\right)^{4}}{1+m_{S}^{4}/m_{N_{1}}^{4}},\label{rd}
\end{equation}
where $\left\langle v_{r}^{2}\right\rangle \simeq6/x_{f}\simeq6/25$
is the thermal average of the relative velocity squared of a pair
of two $N_{1}$ particles, $M_{pl}$ is the Planck mass, and $g_{\ast}\left(T_{f}\right)$
is the total number of effective massless degrees of freedom at the
freeze-out.

\begin{figure}[htp]
\begin{centering}
\includegraphics[width=0.55\textwidth]{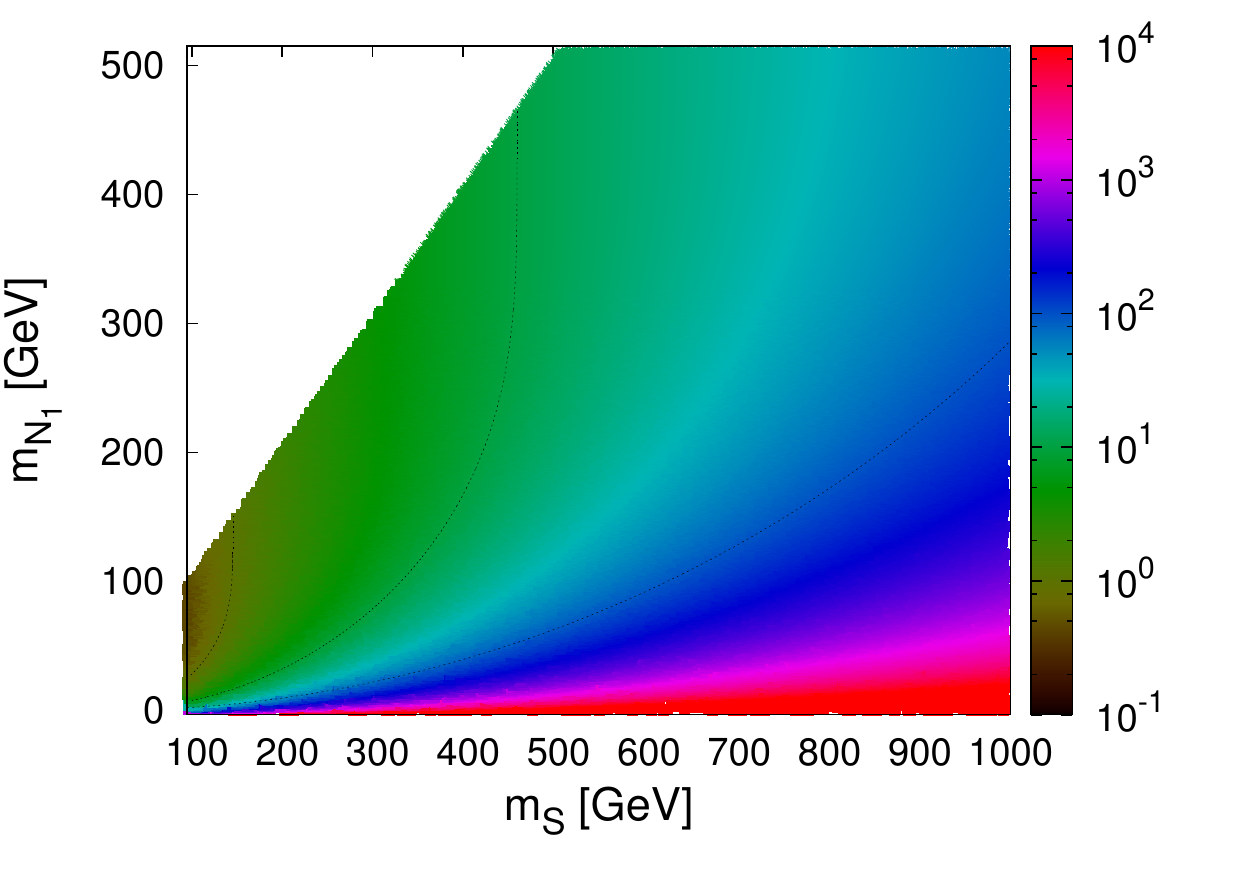} 
\par\end{centering}
\caption{Dark matter mass versus the charged scalar mass, where in the palette one reads the coupling combination
$\sum_{\alpha\beta}\left|g_{1\alpha}g^{*}_{1\beta}\right|^{2}$, that enters in the expression of the relic density and 
can affect the rates of the LFV processes. The dashed curves (from left to right) represent the values 
$\sum_{\alpha\beta}\left|g_{1\alpha}g^{*}_{1\beta}\right|^{2}=1,~10,~100$, respectively.}
\label{r-d} 
\end{figure}

To see the impact of the DM relic density on the model parameters,
we present in Fig.~\ref{r-d} the quantity $\sum_{\alpha\beta}\left|g_{1\alpha}g_{1\beta}^{*}\right|^{2}$
as a palette in the contours $m_{N_{1}}$ versus $m_{S}$ within the conditions $m_{N_{1}}<m_{S}$
and $m_{S}>100$ GeV being imposed. For values of $\sum_{\alpha\beta}\left|g_{1\alpha}g_{1\beta}^{*}\right|^{2}$ 
larger than 10 (corresponding to the light-greenish blue color
in the palette), it is difficult to maintain all LFV process branching
ratios within the current experimental limits, and an extreme fine-tuning is required. On the other hand,
the heavier $N_{1}$ is, the more restricted the allowed range
of the charged scalar mass is. Therefore, the most viable range of the
masses that are consistent with both the DM relic density and the current
bounds on LFV processes are $m_{N_{1}}<200~\text{GeV}$ and $m_{S}<300~\text{GeV}$
while keeping $m_{N_{1}} < m_{S}$.

With regard to the constraint from the DM direct detection experiments, since the interactions of $N_1$ involve 
only a charged lepton, the DM-nucleus scattering is absent at the tree level\footnote{If the mass of $N_1$ arises from 
the vacuum expectation value of some singlet scalar field, then a low-energy effective operator of the form 
$\bar{q} q N_1 N_1$ will be generated at the tree level (see the last reference in Ref.~\cite{AMN}).}, and cannot be 
induced at one loop via the exchange of a photon, since for a Majorana particle the magnetic moment vanishes 
identically. But if the next lightest RH neutrino, say, $N_2$, is quasidegenerate with a mass splitting of the order 
or less than a $\text{few keV}$, then the inelastic scattering $N_1 + \text{nuclues} \rightarrow N_2 + \text{nucleus}$ 
is possible. However, such a situation is highly unnatural for such a tiny mass splitting to be stable under a 
radiative correction which can render the scattering kinematically forbidden.

\section{Constraints from LEP-II}

\begin{figure}[htp]
\begin{centering}
\includegraphics[width=0.48\textwidth,height=0.225\textheight]{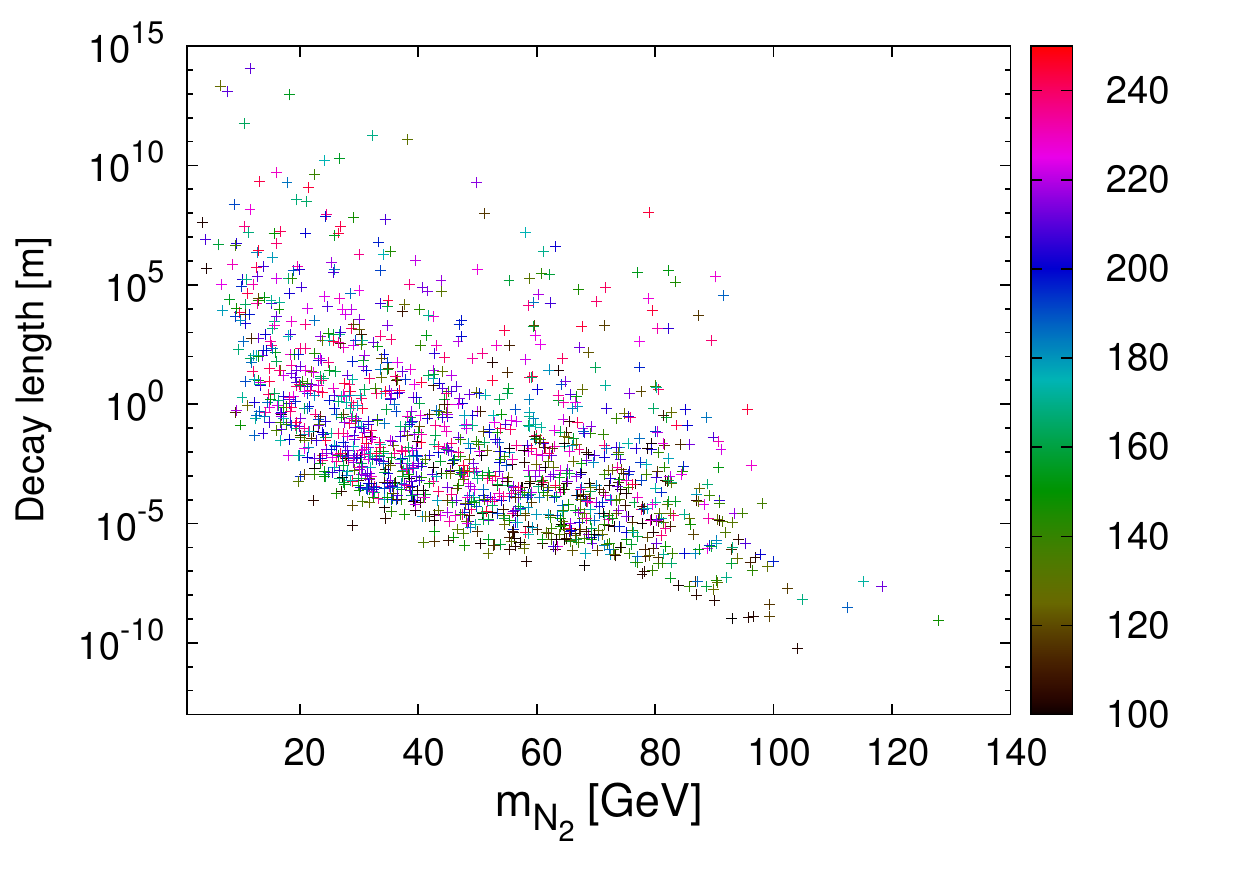}~\includegraphics[width=0.48\textwidth,height=0.225\textheight]{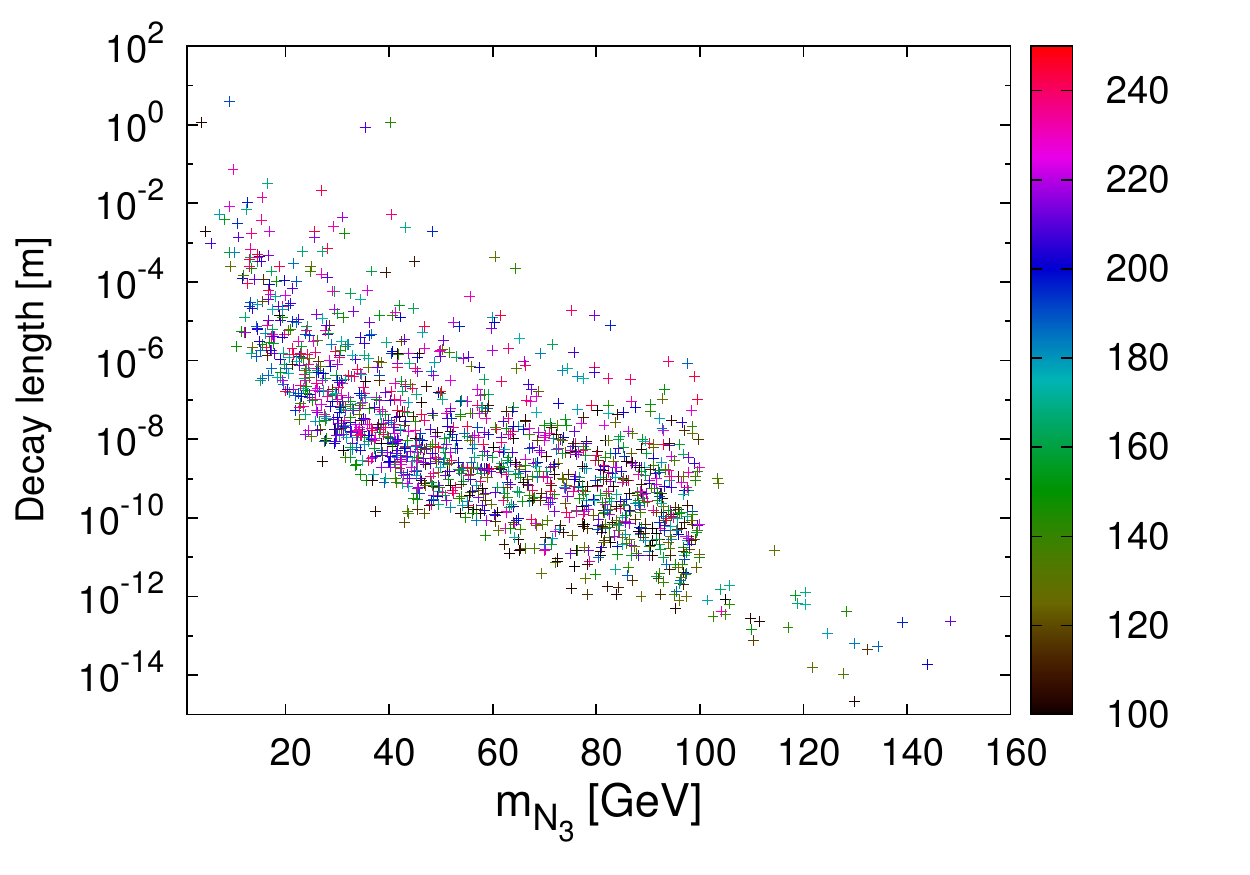} 
\par\end{centering}
\caption{The decay length of the RH neutrinos $N_2$ (left) and $N_3$ (right) as a function of $m_{N_2}$ and $m_{N_3}$, respectively. 
The palette represents the charged scalar mass $m_S$ [GeV].}
\label{decl} 
\end{figure}

Here we consider the analysis of single- and multiphoton events with missing energy by the L3 detector at LEP, 
for center-of-mass energies between 189 and 209 GeV. It was found that the cross section of the process 
$e^{+}e^{-}\rightarrow\nu\bar{\nu}\gamma(\gamma)$ is in agreement with the SM expectations, and there was no evidence
for a massive neutral particle with a significance higher than three that can be produced at LEP-II. This result can 
have a significant impact on the parameter space relevant for DM and neutrino oscillation data. Thus, we
confront thousands of randomly generated benchmark points that respect the different DM and LFV
constraints together with the LEP-II data. As a subsequent study of the electron-positron (electron-electron) 
collision on the International Linear Collider (ILC) will be carried out in the next sections, it is necessary to sort out the benchmark points of 
$ N_2 $ and $ N_3 $ based on whether their decay via a three-body process will occur inside or outside the detector. 
In Fig.~\ref{decl}, we present the decay length for $N_2$ and $N_3$, where one can see that $N_3$ decay mostly inside 
the detector, whereas an appreciable fraction of $N_2$ events escape from the detector. Consequently, only $N_1$ and 
the $N_2$ events that decay outside the detector will be counted as missing energy and, hence, are subject to the LEP 
constraint. In all our analyses, we apply this selection criterion.

We consider the center-of-mass energies $\sqrt{s}=188.6~\text{GeV}$ and $\sqrt{s}=207~\text{GeV}$ at the highest integrated luminosities
$176$ and $130.2~\text{pb}^{-1}$, respectively. In order to increase the signal significance, 
we apply the following kinematical cuts used by the L3 Collaboration to look for a high-energy single photon~\cite{LEP}:

\begin{itemize}
\item [$\bullet $] the polar angle of the photon: $\left|\text{cos}\,\theta_{\gamma}\right|<0.97,$%\\
%\vspace{-0.3cm}
\item [$\bullet$] the transverse momentum of the photon: $p_{t}^{\gamma}>0.02\sqrt{s},$ and%\\
\item [$\bullet$] the energy of the photon: $E_{\gamma}>1\,\text{GeV}$.
\end{itemize}

Using the L{\footnotesize{AN}}HEP/C{\footnotesize{CALC}}HEP packages~\cite{LanHEP,CalcHEP}, to which the interaction terms in (\ref{LL}) 
are implemented, we compute the cross sections of the signal $e^{-}e^{+}\rightarrow\gamma+E_{miss}$ and the background
$e^{-}e^{+}\rightarrow\nu_{i}\bar{\nu}_{j}\gamma$ for the aforementioned benchmark points. Furthermore, we evaluate the 
significance at the corresponding luminosity as a function of the dark matter mass. Moreover, we found that the cross section is sensitive to the inverse of $m_S m_{N_i}$ as well products of the interaction couplings, and hence we present a combination of these parameters. This lead us to derive an exclusion bound on a combination of these parameters.

%is inversely proportional to both $m_S$ and $m_{N_i}$ and involves products of the couplings $g$'s. This lead us to derive an exclusion bound on a combination of these parameter,
%\textbf{Since the cross section of 
%$e^{-}e^{+}\rightarrow\gamma+N_iN_i$ is inversely proportional to both $m_S$ and $m_{N_i}$ and directly 
%proportional to $|g_{ie}|^4$, we show the quantity $\sum_{i,k}|g_{ie}g_{ke}|^2\left[150~GeV/m_{S}\right]\left[50~GeV/\sqrt{m_{N_i}m_{N_k}}\right]$ in the 
%palette in Fig.~\ref{lep}.}

\begin{figure}[htp]
\begin{centering}
\includegraphics[width=0.48\textwidth,height=0.225\textheight]{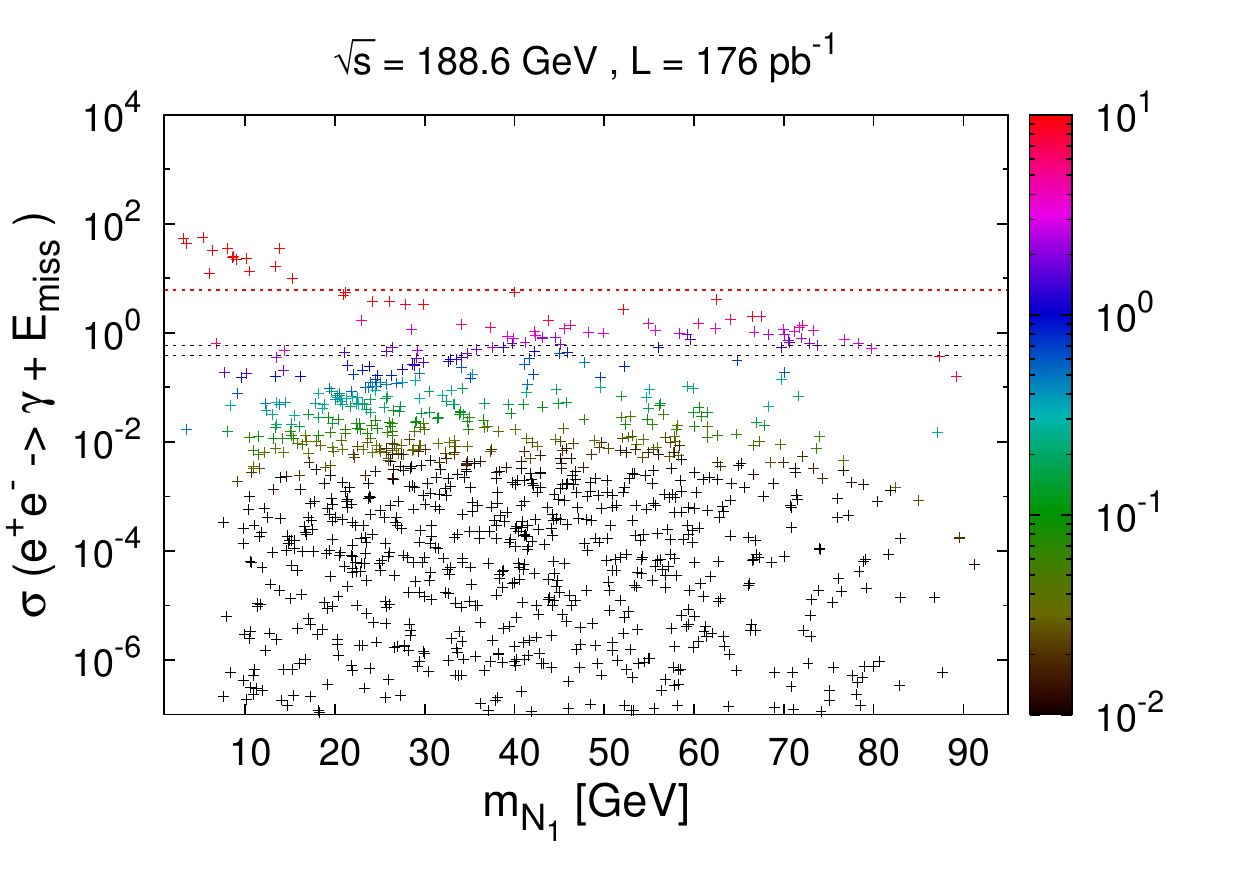}~\includegraphics[width=0.48\textwidth,height=0.225\textheight]{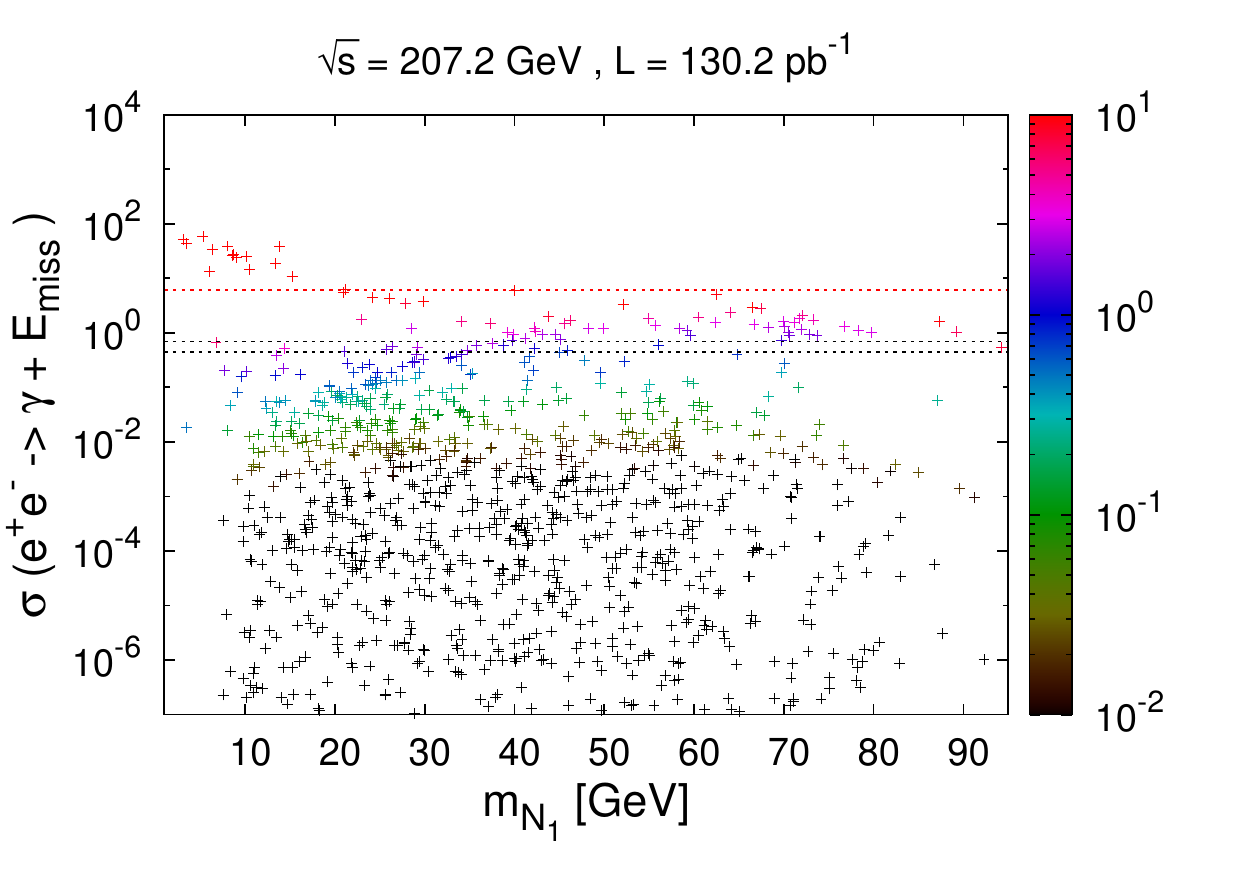} 
\par\end{centering}
\caption{The cross section for the randomly chosen benchmark points for the process $e^{-}e^{+}\rightarrow\gamma+E_{miss}$ 
at LEP as a function of $m_{N_{1}}$ for the CM energies $\sqrt{s}=188.6~\text{GeV}$ (right) and $\sqrt{s}=207.2~\text{GeV}$ 
(left). The palette represents the combination $\Delta$, and the black dashed lines correspond to $S=2,~3$, respectively. The red dashed line corresponds to the background.}
\label{lep} 
\end{figure}

The results are shown in Fig.~\ref{lep}, requiring that at such energies and the corresponding luminosity 
values the significance $S$ must be smaller than three ($S<2$ for a conservative choice) leads to a constraint 
on the model parameters as\footnote{For $S < 2$ the constraint (\ref{LEP}) becomes 
$\sum_{i,k} \left|g_{ie}g_{ke}^{*}\right|^{2} \left[\frac{150~\text{GeV}}{m_{S}}\right] \left[\frac{50~\text{GeV}}{\sqrt{m_{N_i}m_{N_k}}}\right] <1.65$.}
\begin{eqnarray}
\Delta =\sum_{i,k} \left|g_{ie}g_{ke}^{*}\right|^{2} \left[\frac{150~\text{GeV}}{m_{S}}\right] \left[\frac{50~\text{GeV}}{\sqrt{m_{N_i}m_{N_k}}}\right] <1.95,\label{LEP}
\end{eqnarray}
where the summation is performed over RH neutrinos that contribute to the missing energy, to which according to Fig.~\ref{decl} only $N_1$ and $N_1$ contribute. From the large values of $\Delta $ in the palette, one concludes that the absence of new physics at LEP can put a significant 
constraint on the strength of the interaction $S^{+}\overline{N_{i}}\ell_{\alpha_{R}}$, especially when the RH 
neutrino is lighter than $50~\text{GeV}$. Consequently, it can have an important impact on the scale of the 
generated neutrino mass in models based on such a type of interaction.

\section{Possible Signatures at Lepton Colliders}

The study of an electron-positron and/or electron-electron collision at lepton colliders such as the ILC~\cite{ilc} represents a new approach for probing/detecting new physics in the tera scale. 
The ILC is designed to cover center-of-mass (CM) energies from $250$ to $500~\text{GeV}$, with the possibility 
to expand it up to 1 TeV and with the option of using polarized beams for both electrons and positrons. Another 
lepton collider which is under development is the Compact Linear Collider (CLIC) which will provide high-luminosity $e^{+} e^{-}$ collisions with CM energy from $380~\text{GeV}$ to $3~\text{TeV}$~\cite{CLIC}. 
Here, in this work, we consider the main processes where the DM particles are in the final states via the 
interactions in (\ref{LL}) and could induce an excess in the event number relative to the SM background.

\begin{figure}[htp]
\begin{centering}
\includegraphics[width=0.6\textwidth]{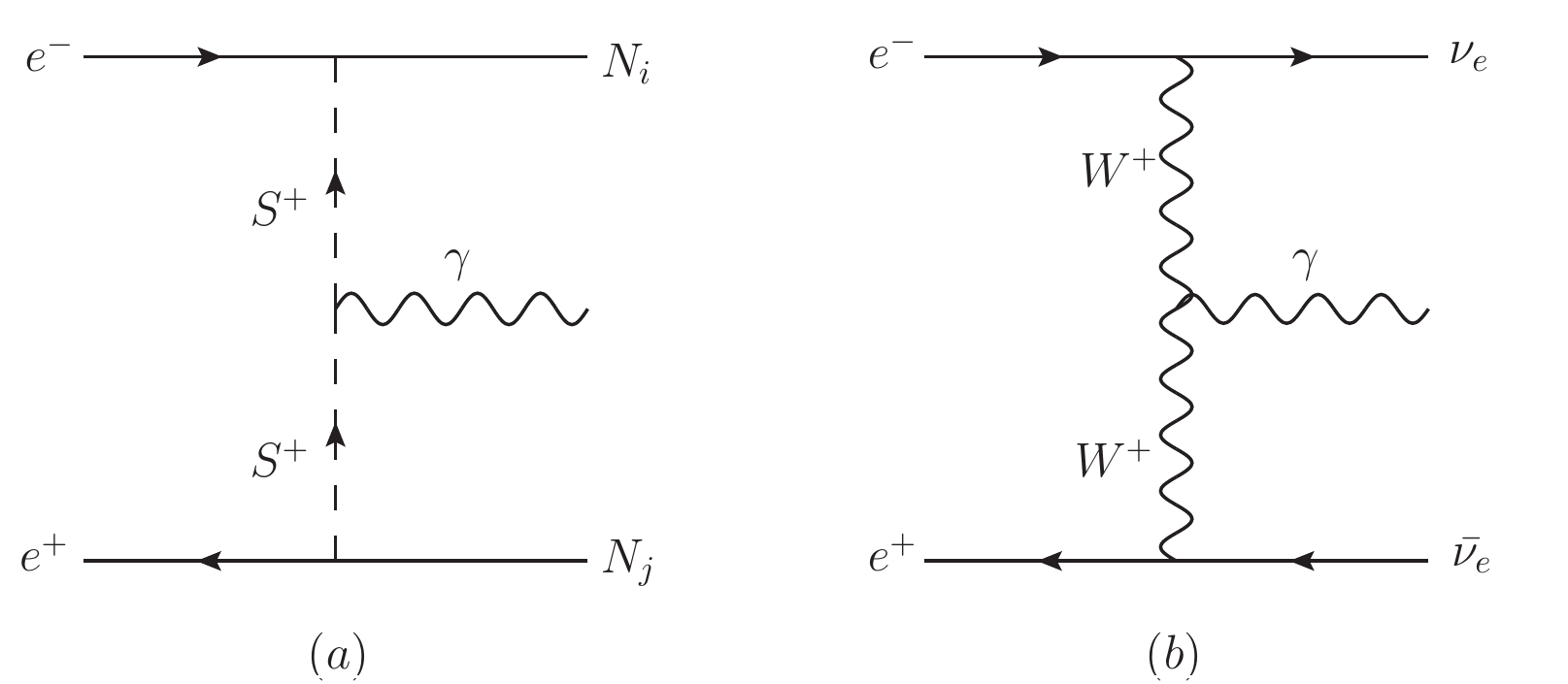}
\par\end{centering}
\caption{The main Feynman diagram (a) [(b)] that contributes to the signal [background] for the process $e^{-}e^{+}\rightarrow\gamma+E_{miss}$.}
\label{diag1} 
\end{figure}
\begin{figure}[htp]
\begin{centering}
\includegraphics[width=0.6\textwidth]{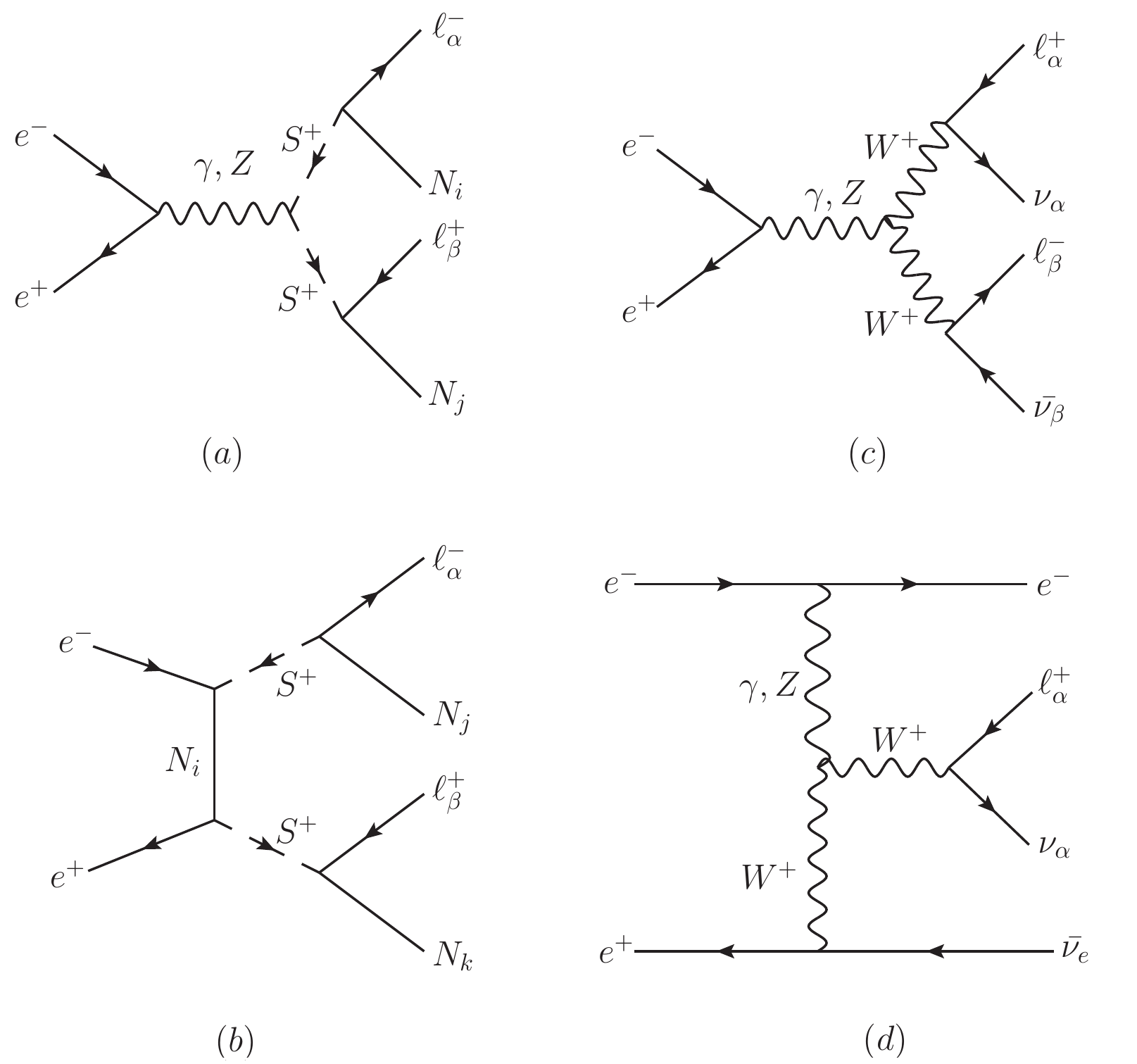}
\par\end{centering}
\caption{The most important Feynman diagrams (a),(b) [(c),(d)] that contribute to the signal 
(background) for the process $e^{-}e^{+}\rightarrow S^{+}S^{-}$.}
\label{diag2} 
\end{figure}
\begin{figure}[htp]
\begin{centering}
\includegraphics[width=0.6\textwidth]{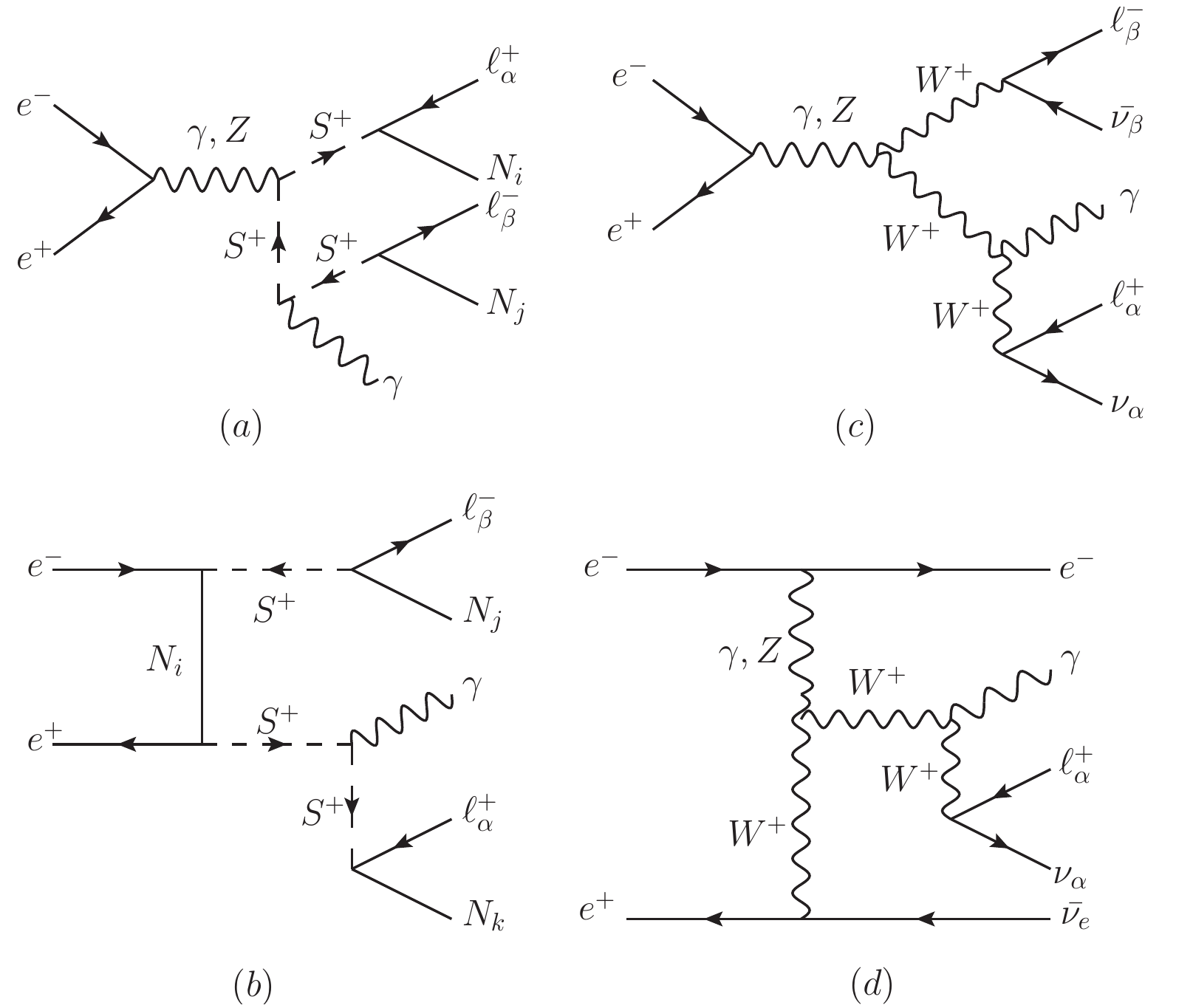} 
\par\end{centering}
\caption{The most important Feynman diagrams (a),(b) [(c),(d)] that contribute to the signal 
(background) for the process $e^{-}e^{+}\rightarrow S^{+}S^{-}+\gamma$.}
\label{diag3} 
\end{figure}
\begin{figure}[htp]
\begin{centering}
\includegraphics[width=0.6\textwidth]{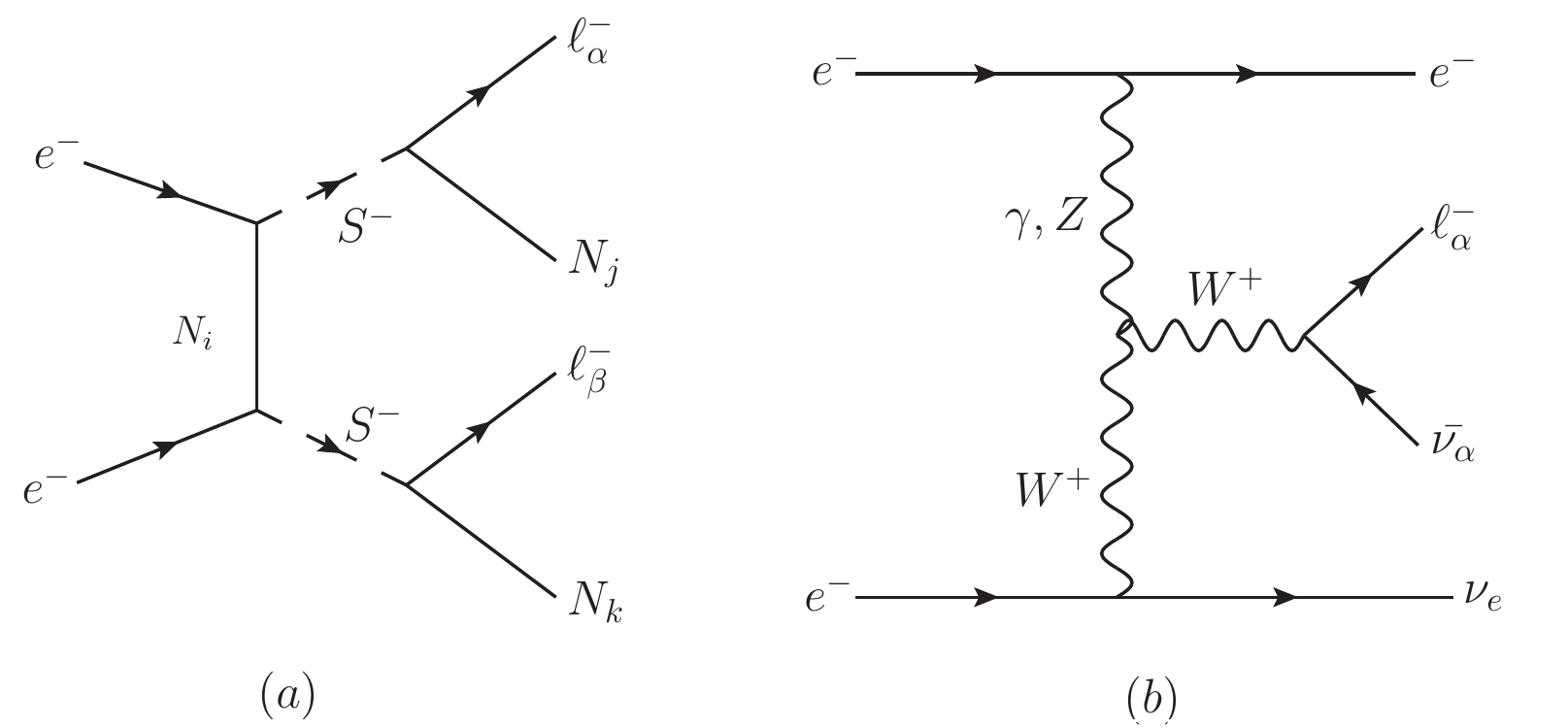}
\par\end{centering}
\caption{The main Feynman diagram (a) [(b)] that contributes to the signal (background) for the process $e^{-}e^{-}\rightarrow S^{-}S^{-}$.}
\label{diag4} 
\end{figure}

\begin{figure}[htp]
\begin{centering}
\includegraphics[width=0.48\textwidth,height=0.225\textheight]{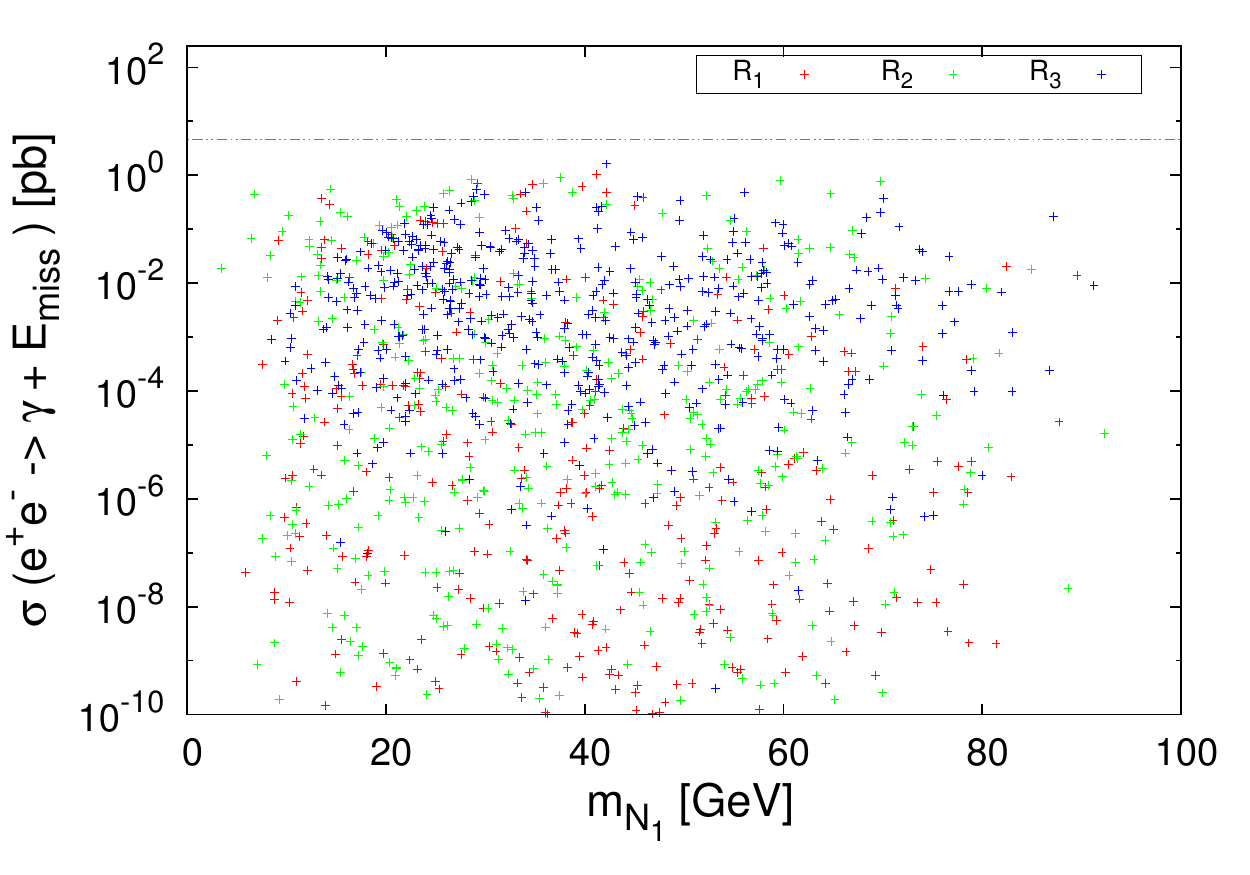}~\includegraphics[width=0.48\textwidth,height=0.225\textheight]{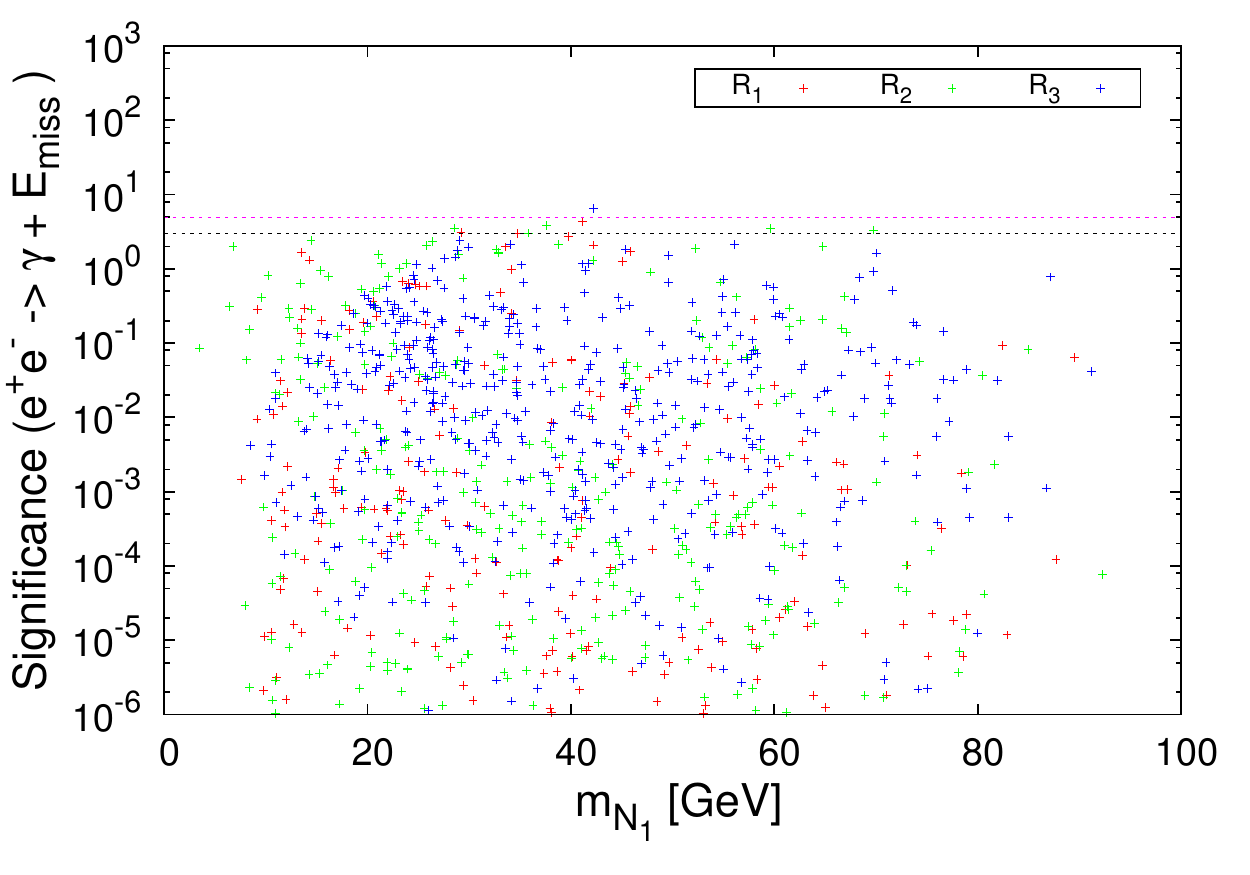}\\
\includegraphics[width=0.48\textwidth,height=0.225\textheight]{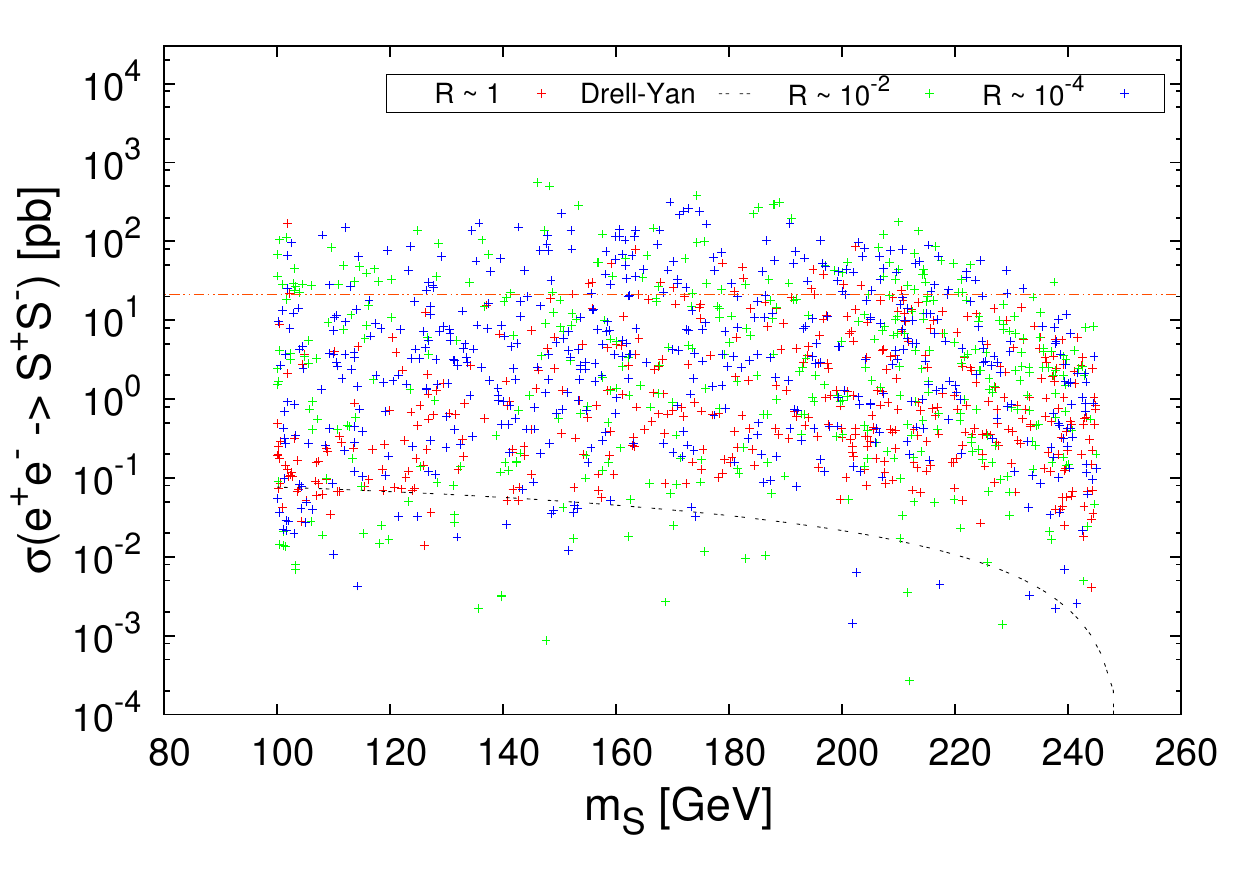}~\includegraphics[width=0.48\textwidth,height=0.225\textheight]{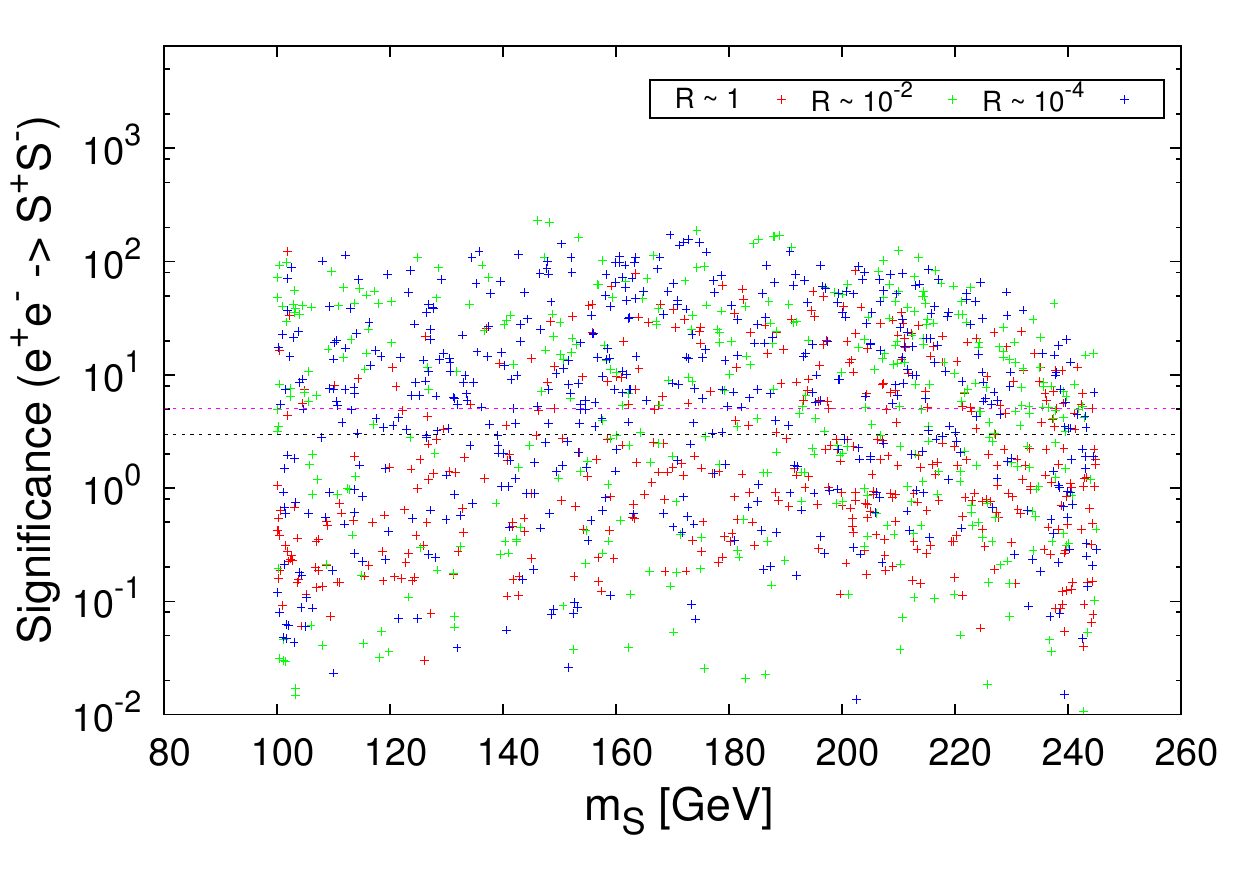}\\
\includegraphics[width=0.48\textwidth,height=0.225\textheight]{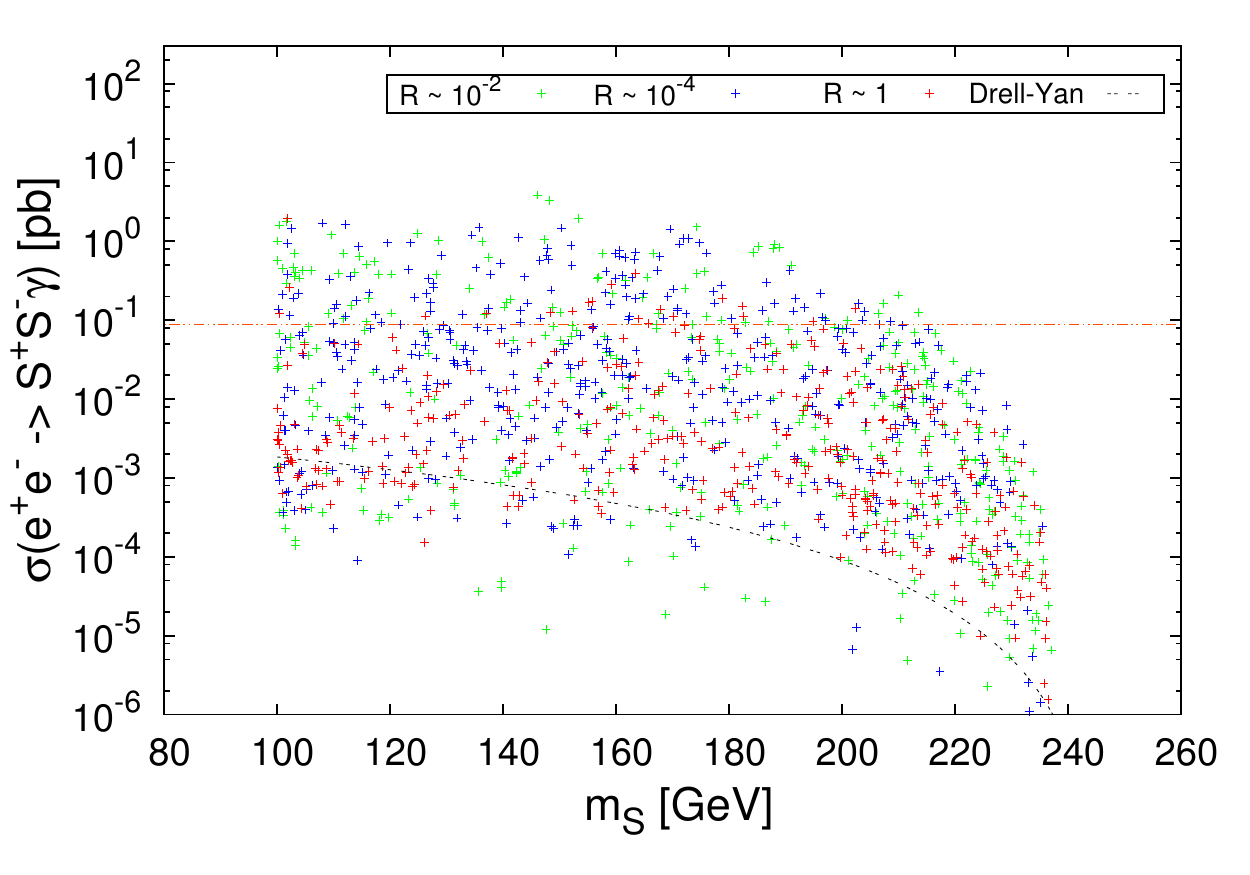}~\includegraphics[width=0.48\textwidth,height=0.225\textheight]{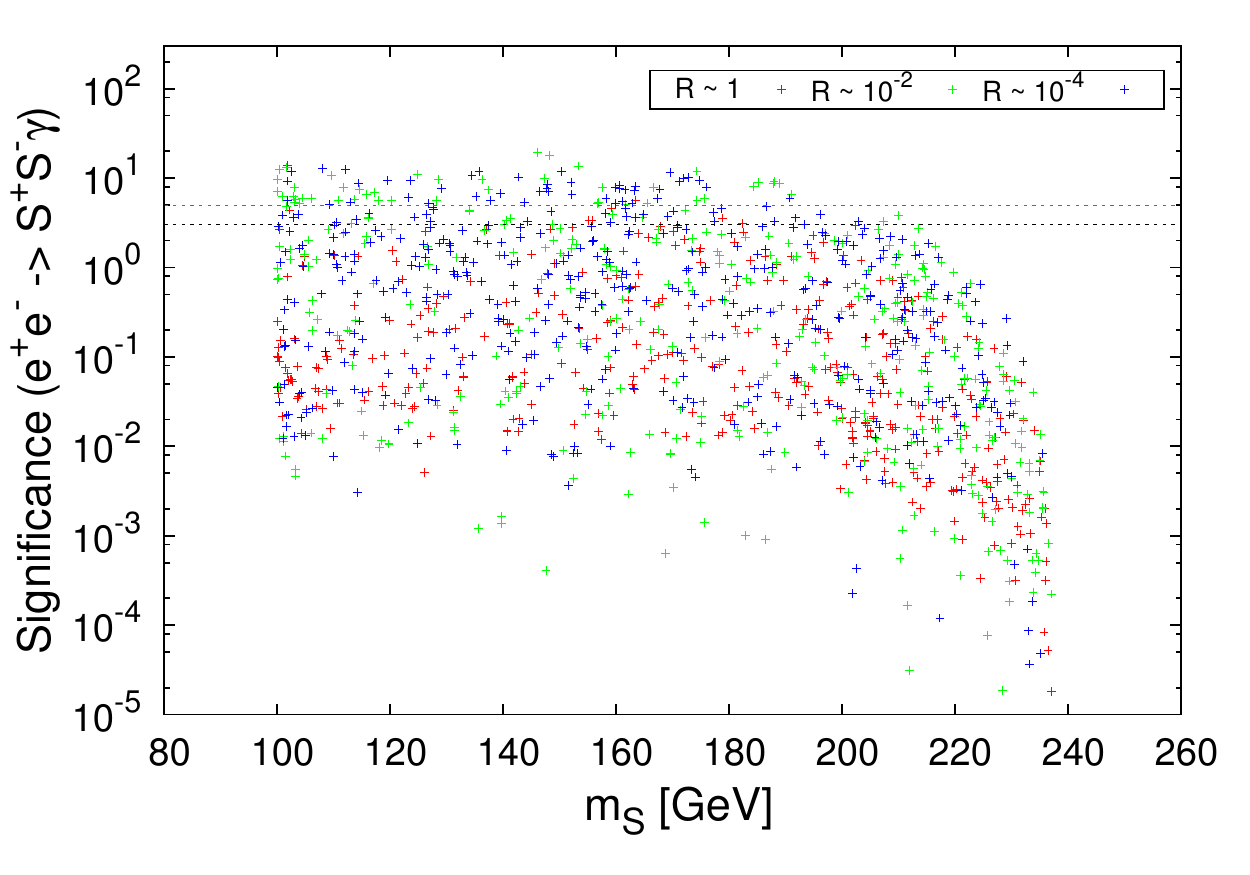}\\
\includegraphics[width=0.48\textwidth,height=0.225\textheight]{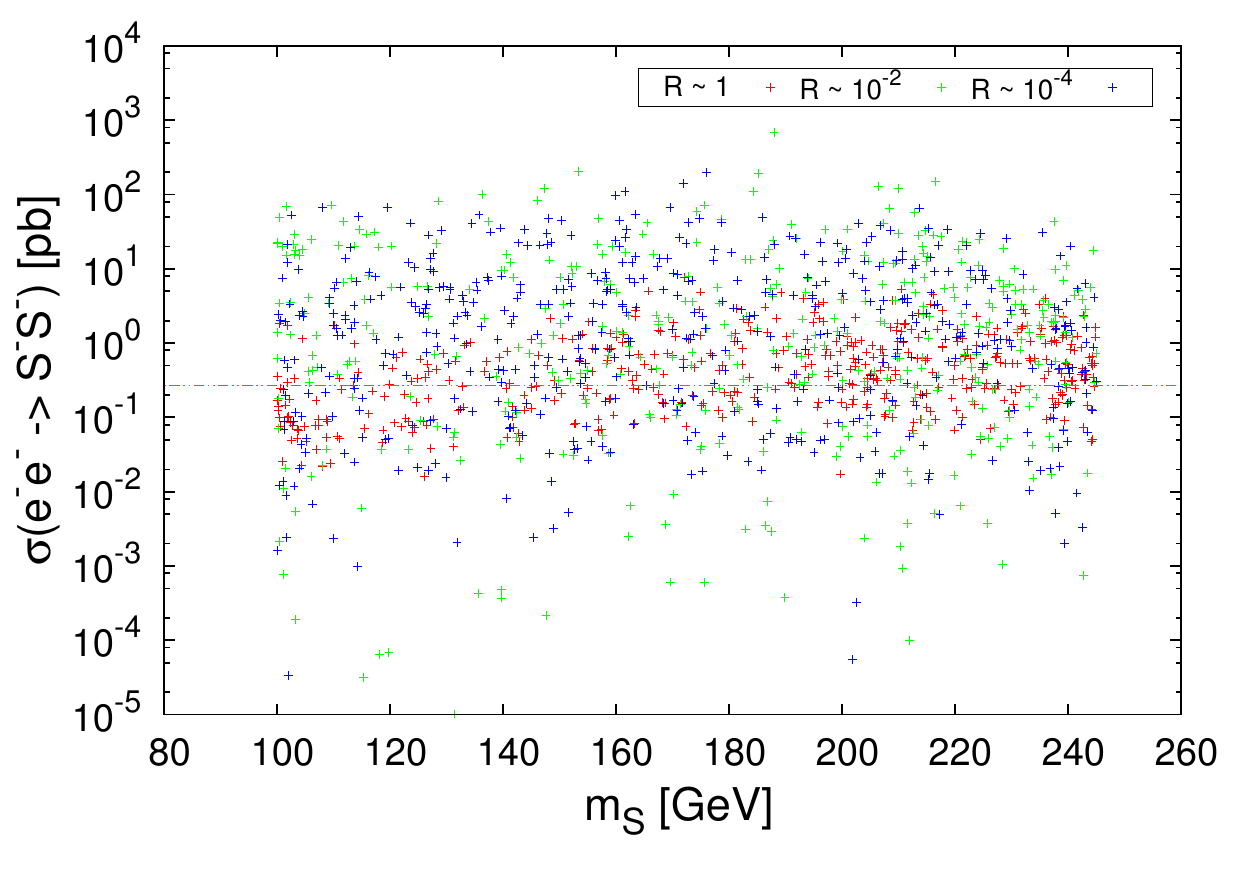}~\includegraphics[width=0.48\textwidth,height=0.225\textheight]{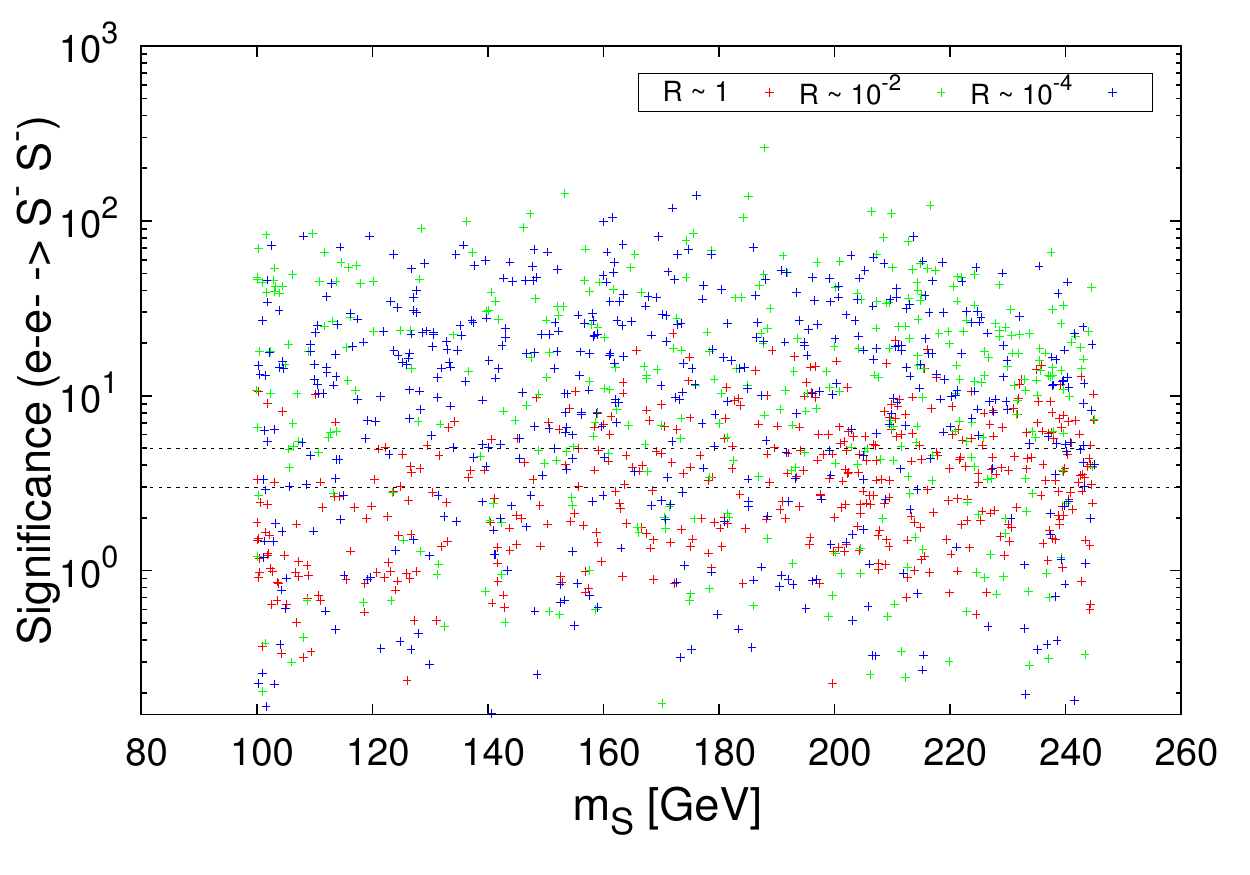} 
\par\end{centering}
\caption{The cross section values (left) and the corresponding
significance values (right) at luminosity $100~\text{pb}^{-1}$ in a function of $m_S$ for randomly chosen sets of parameters
 within three specific values of the fine-tuning parameter $R$.
The red lines (left) represent the background value, and the dashed one represents
the Drell-Yan contribution, and the dashed
lines (right) represent $S=3,5$, respectively.}
\label{CS} 
\end{figure}

The most interesting signature involves a pair of DM in the final state with one (or more) photon(s) 
irradiated from the intermediate charged scalar as depicted in Fig.~\ref{diag1}. In this case, the background 
corresponds to a single (or multiple) photon(s) plus light neutrinos which can be reduced by applying the 
cut $E_{\gamma}>8\,\text{GeV}$ over the energy of the photon. Another potential signature is a pair production 
of charged scalars $S^{+} S^{-}$ without or with a photon in the final state, as shown in Figs.~\ref{diag2} and \ref{diag3}, where $S^{\pm}$ decays into a RH neutrino and a charged lepton. In this case, the background 
contributing to the signal comes from the process $e^{+}e^{-}\rightarrow W^{+}W^{-}$ where each $W$ decays into 
a charged lepton and a light neutrino. Also, a same-sign pair of charged scalars can be produced in electron-electron 
collisions, as seen in Fig.~\ref{diag4}, where a final state with same-sign leptons with missing energy can 
be observed. Therefore, in this study we consider the following processes:
 \begin{eqnarray}
e^{-}e^{+}&\rightarrow &\gamma+E_{miss}, \nonumber\\
e^{-}e^{+} &\rightarrow & S^{+}S^{-}\rightarrow\ell_{\alpha}^{+}\ell_{\beta}^{-}+E_{miss},\nonumber\\
e^{-}e^{-} &\rightarrow & S^{-}S^{-}\rightarrow\ell_{\alpha}^{-}\ell_{\beta}^{-}+E_{miss}\nonumber\\
e^{-}e^{+}&\rightarrow &\gamma+S^{+}S^{-}\rightarrow\gamma+\ell_{\alpha}^{+}\ell_{\beta}^{-}+E_{miss},\label{signal}
\end{eqnarray} 
and limit ourselves to a center-of-mass energy of $\sqrt{s}=500~\text{GeV}$ and with a luminosity $L=100~\text{pb}^{-1}$. 
We generate three sets of benchmark points according to different values of the ratio $ R_1 \approx 1$, $ R_2 \approx 10^{-2}$,
and $ R_3 \approx 10^{-4}$ which are in agreement with the observed DM relic density and respect the LFV constraints. 
The background processes (\ref{signal}) are due to the exchange of $W/Z/\gamma$ gauge bosons, and the corresponding Feynman 
diagrams can be similar to the ones for the signal. 
In Fig.~\ref{CS}, we show the cross section values and the corresponding significance at $L=100~\text{pb}^{-1}$ for the 
processes (\ref{signal}) as a function of the charged scalar mass before applying any cut.\footnote{Except the cut 
$E_{\gamma}>$ 8 GeV and $|\text{cos}~\theta_{\gamma}|<0.998$ for the monophoton and $S^{-}S^{+}+\gamma$ channels.}

%One remarks that the cross section values of the processes (\ref{signal}) in Fig.~\ref{CS} varies over seven orders of magnitudes as its sensitively depends on our choice of the parameters space. 
It is worth noting that $\sigma (e^{-} e^{+} \rightarrow S^{+} S^{-})$ and 
$\sigma (e^{-} e^{+} \rightarrow S^{+} S^{-} + \gamma)$ are dominated by the diagrams due to interactions 
(\ref{LL}) rather than the Drell-Yan diagrams. The interference contributions could be 
negative and can make the cross section smaller than the Drell-Yan one, as shown in Fig.~\ref{CS} (left). 
As one can see, $\sigma (e^{-} e^{+} \rightarrow S^{+} S^{-} + \gamma)$ is about $100$ times smaller than $\sigma (e^{-} e^{+} \rightarrow S^{+} S^{-})$ 
due to the $e^2/{4\pi}$ suppression from the coupling of the charged scalar to the photon. The 
production cross section of the same-sign charged scalars via an electron-electron collision is huge compared to the 
background, and hence the signal significance in this case could be large even for low luminosity. Therefore, this 
process is a clean and direct probe for RH neutrinos at the ILC.

\section{Benchmark Analysis}

In this section, we consider three benchmark points with fine-tuning parameters $ R_1 \approx 1,~ R_2 \approx 10^{-2}$, and 
$ R_3 \approx 10^{-4}$, respectively.
As can be seen from Table-~\ref{tab-point}, the choice of the ratios $R_i$ limits substantially our freedom of the model 
parameter space. Using C{\footnotesize{CALC}}HEP~\cite{CalcHEP}, we generate the distributions for different kinematic variables for both the signal and background of the processes $e^{-}e^{+}\rightarrow \gamma+E_{miss}$,\ $e^{-}e^{+}\rightarrow S^{-}S^{+}$,\ 
and $e^{-}e^{+}\rightarrow S^{-}S^{+}+\gamma$ at $500~\text{GeV}$ center-of-mass energy. Then, we apply the kinematical 
cuts that optimize the signal detection over the background and estimate the corresponding signal significance for each process.

\begin{table}[htp]
\begin{centering}
\begin{adjustbox}{max width=\textwidth}
\begin{tabular}{cccc}
\hline 
Point & $B_{1}\left(R_{1}\right)$ & $B_{2}\left(R_{2}\right)$ & $B_{3}\left(R_{3}\right)$\tabularnewline
\hline 
$g_{1e}$ & $(7.506+i0.014)\times10^{-1}$ & $(1.8284+i0.103)$ & $(-0.103+i0.201)$ \tabularnewline
\hline 
$g_{2e}$ & $(-0.26819-i1.5758)\times10^{-4}$ & $(1.543+i3.004)\times10^{-4}$ & $(0.654-i2.616)\times10^{-2}$\tabularnewline
\hline 
$g_{3e}$ & $(-1.360-i0.707)$ & $(0.313-i0.549)$ & $(-0.869-i0.878)$\tabularnewline
\hline 
$m_{S}$(GeV) & $196.75$ & $242.81$ & $104.47$\tabularnewline
\hline 
$m_{N_{1}}(\text{GeV})$ & $25.788$ & $43.764$ & $38.306$\tabularnewline
\hline 
$m_{N_{2}}(\text{GeV})$ & $28.885$ & $58.182$ & $56.481$\tabularnewline
\hline 
$m_{N_{3}}(\text{GeV})$ & $36.274$ & $67.511$ & $72.440$\tabularnewline
\hline 
\end{tabular}\end{adjustbox} 
\par\end{centering}
\caption{Three benchmark points selected from the parameter space of the model
for a detailed analysis.}
\label{tab-point} 
\end{table}

\subsection{The monophoton final state $\gamma+E_{miss}$}

First we use the distributions with precuts $E_{\gamma}>8\,\text{GeV}$ and $|\text{cos}\,\theta_{\gamma}|<0.998$ 
and $E_{miss}>100~\text{GeV}$ and then deduce the cuts that reduce the contribution of the background relative 
to the signal. We find this can be achieved for the following cuts : 
\begin{eqnarray}
 8~\text{GeV}<E_{\gamma}<300~\text{GeV},~~~~~~~|\text{cos}~\theta_{\gamma}|<0.998,~~~~~~~E_{miss}>300~\text{GeV}.\label{cut-monoA}
\end{eqnarray}

\begin{figure}[htp]
\begin{centering}
\includegraphics[width=0.48\textwidth]{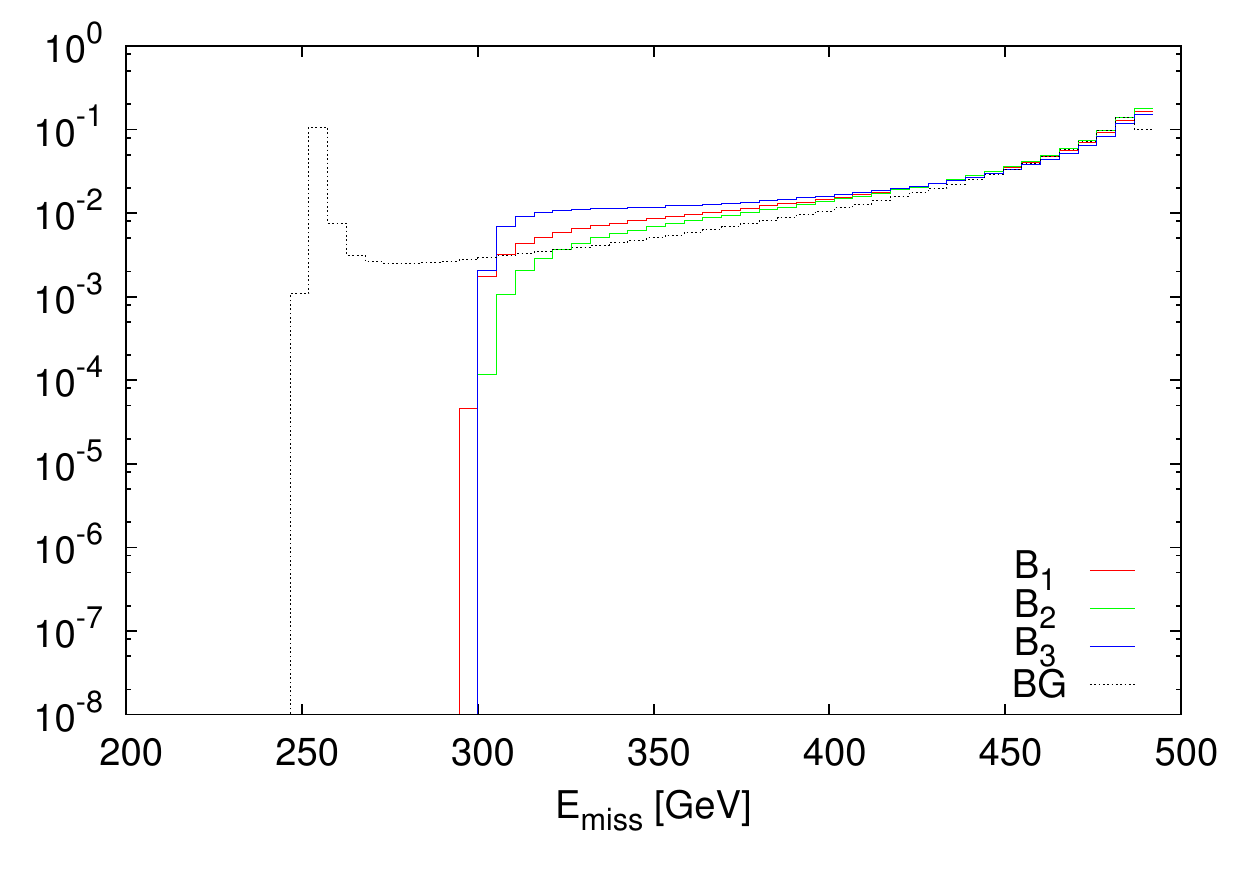}~\includegraphics[width=0.48\textwidth]{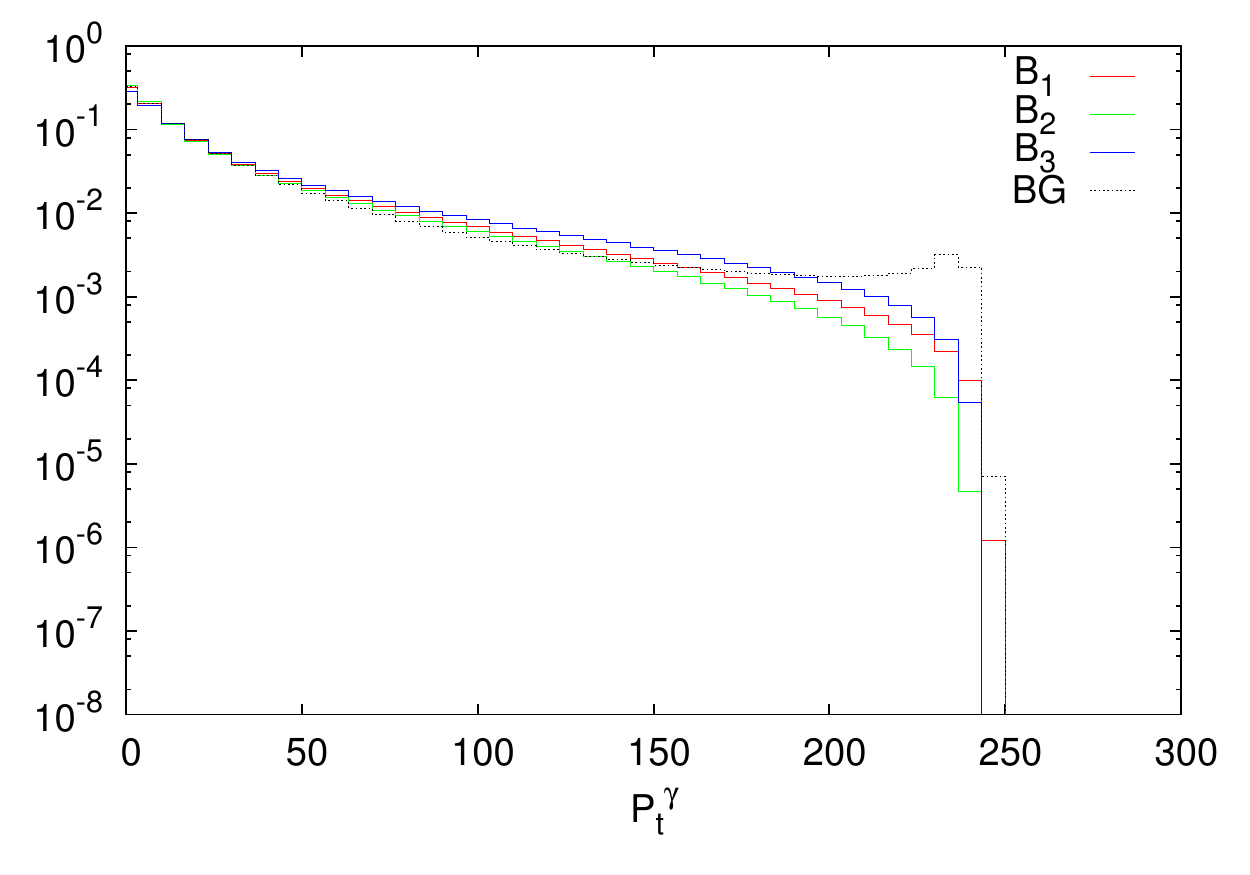} 
\par\end{centering}
\caption{The normalized distributions $E_{miss}$ and $p_{t}^{\gamma}$ of the process $e^{-}e^{-}\rightarrow\gamma+E_{miss}$
at $\sqrt{s}=500$ GeV. The cuts used here are $E_{\gamma}>8\,\text{GeV}$
and $|\textbf{cos}\,\theta_{\gamma}|<0.998$.}
\label{distr} 
\end{figure}

The distributions of the missing energy and the photon transverse momentum variables after applying the above cuts 
are shown in Fig.~\ref{distr}.

\begin{figure}[htp]
\begin{centering}
\includegraphics[width=0.55\textwidth]{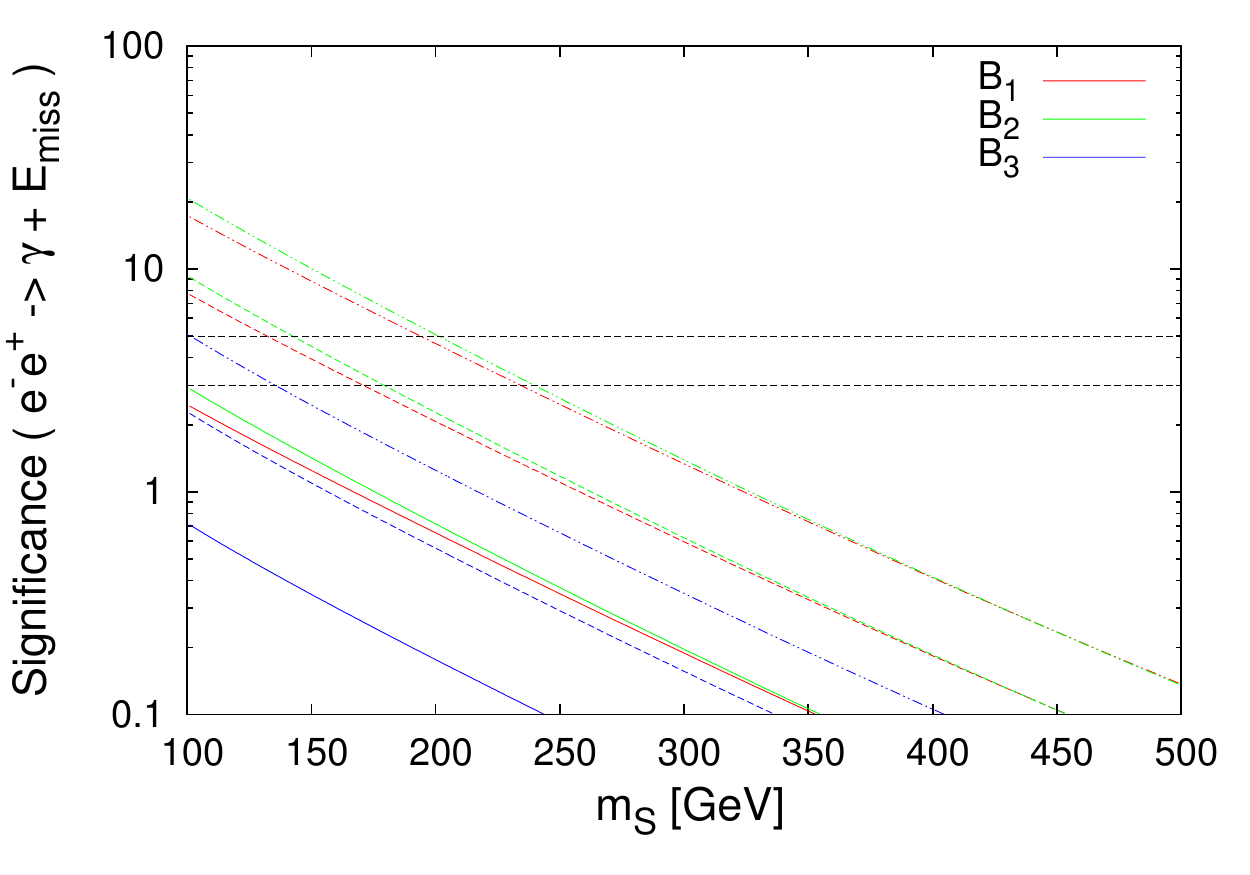} 
\par\end{centering}
\caption{The signal significance for the process $e^{-}e^{+}\rightarrow\gamma+E_{miss}$ as a function 
of $m_{S}$ for the chosen value of $g_{i\alpha}$ given in Table-~\ref{tab-point}
at an integrated luminosity of 10 (solid lines), 100 (dashed lines), and $500~\text{fb}^{-1}$ 
(dash-dotted lines). The horizontal dashed lines correspond to a 3 and 5 sigma significance.}
\label{s-c} 
\end{figure}

By varying the charged scalar mass, we show in Fig.~\ref{s-c} the signal significance at integrated 
luminosities $L = 10, 100,$ and $500~\text{fb}^{-1}$ for each benchmark point.

{As can be seen, for charge scalars lighter than $200~\text{GeV}$, the signal-to-background ratio could be larger 
than unity, which makes the RH neutrino signal detectable at the ILC for a luminosity of a few hundred 
$\text{fb}^{-1}$ whatever the fine-tuning parameter value}. However, for charged scalars heavier than $300~\text{GeV}$, it requires a very high 
luminosity for the signal to be detected with a 5 sigma significance or larger.

\subsection{Final state $S^{+}S^{-}(\gamma)$}

\begin{figure}[htp]
\begin{centering}
\includegraphics[width=0.33\textwidth,height=0.18\textheight]{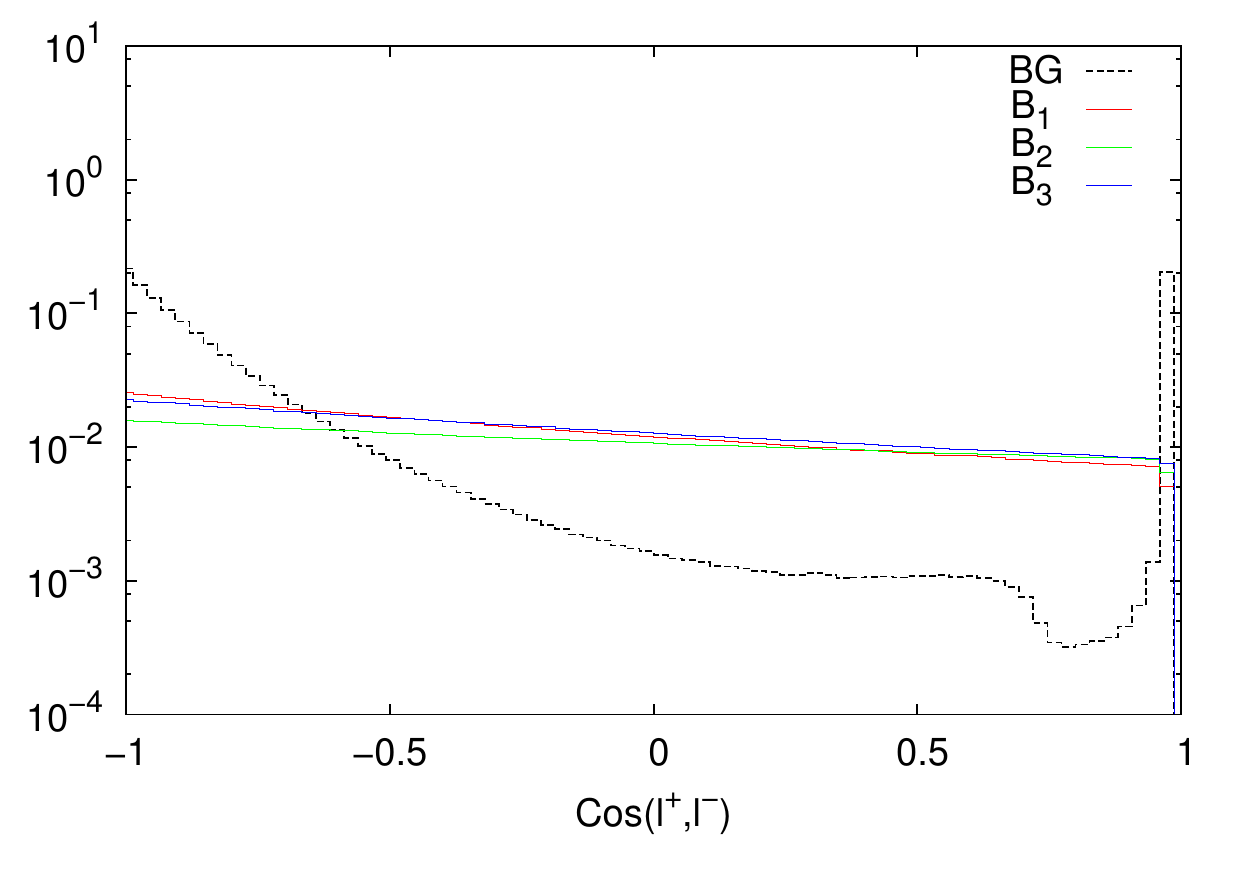}~\includegraphics[width=0.33\textwidth,height=0.18\textheight]{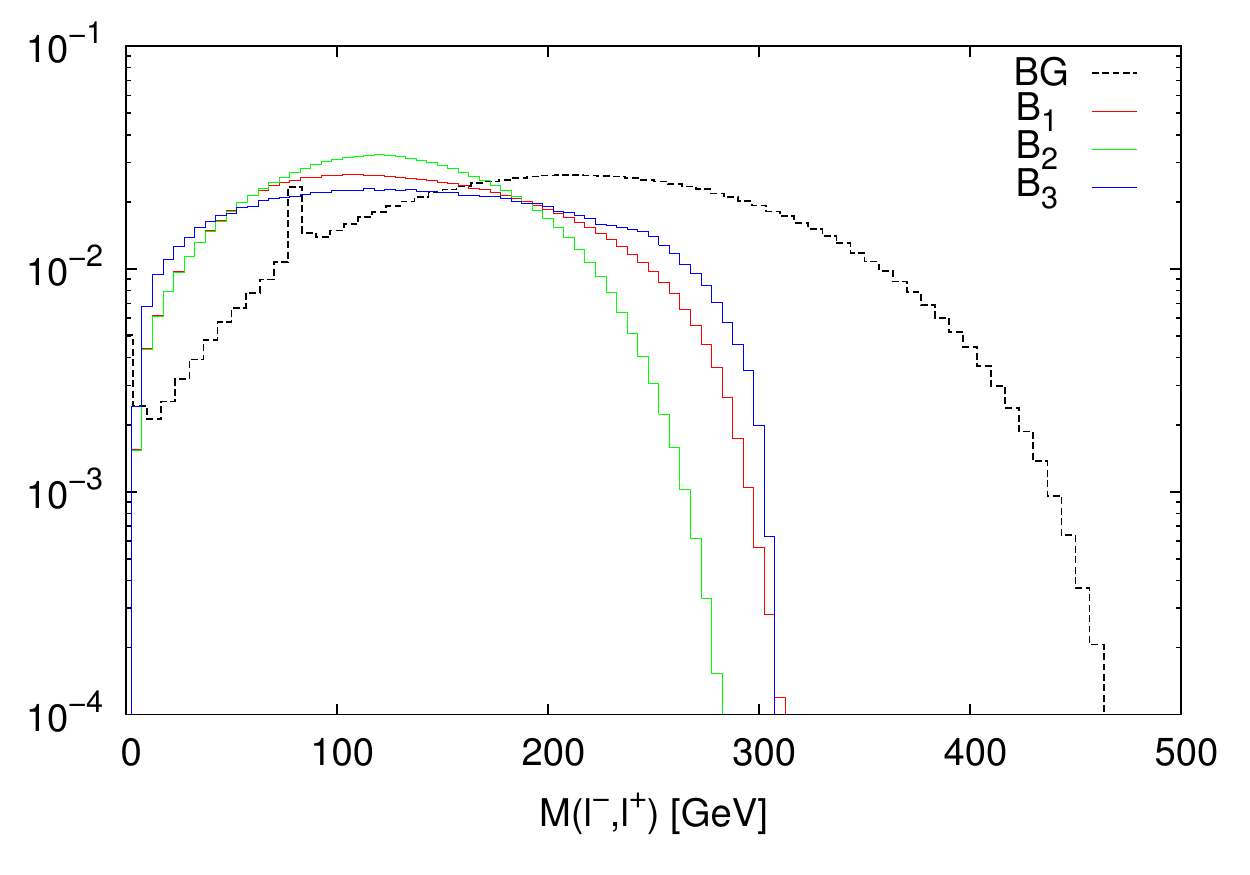}~\includegraphics[width=0.33\textwidth,height=0.18\textheight]{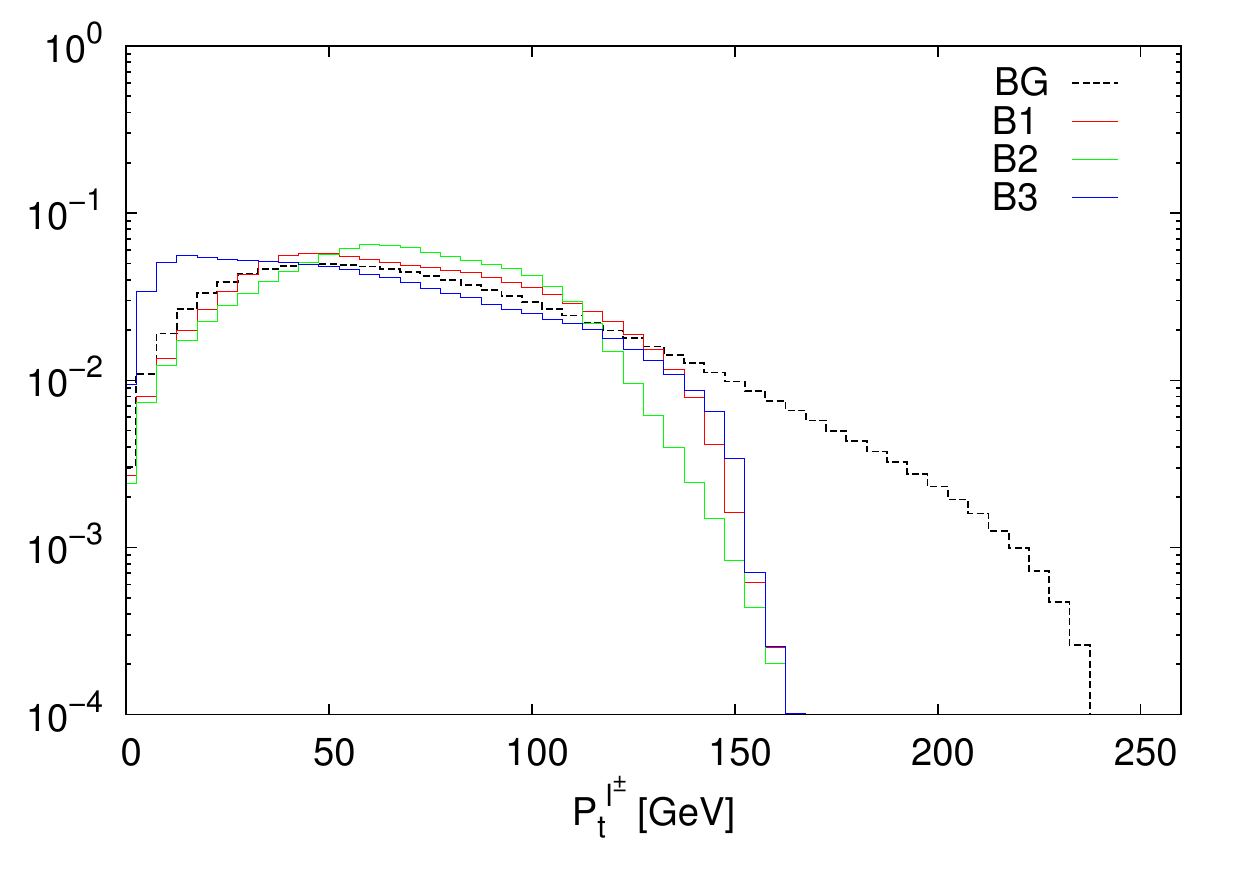} 
\par\end{centering}
\caption{The angular, invariant mass, and transverse momentum normalized distributions for charged leptons
of process $e^{-}e^{+}\rightarrow S^{-}S^{+}$ at $\sqrt{s}=500~\text{GeV}$.}
\label{distr-eES} 
\end{figure}

\begin{figure}[htp]
\begin{centering}
\includegraphics[width=0.33\textwidth,height=0.18\textheight]{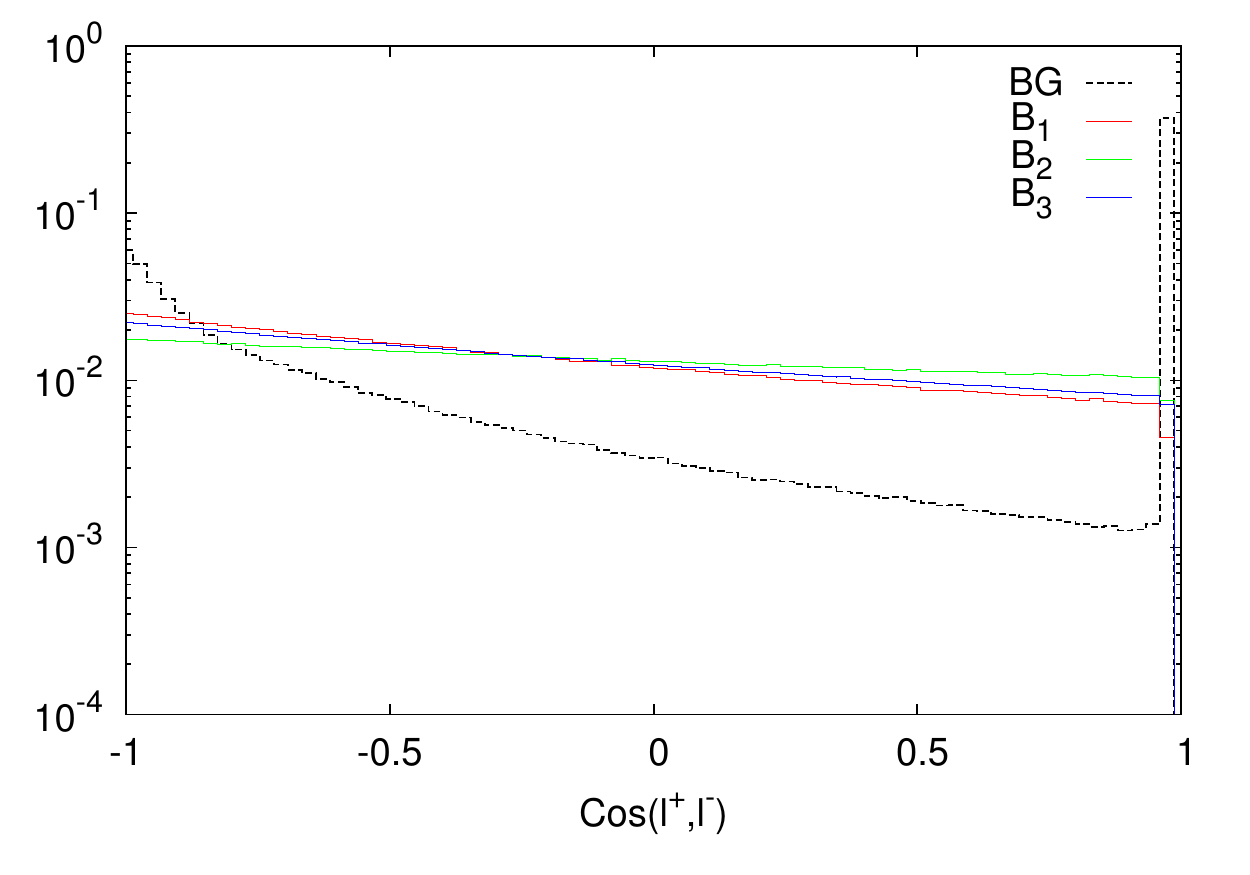}~\includegraphics[width=0.33\textwidth,height=0.18\textheight]{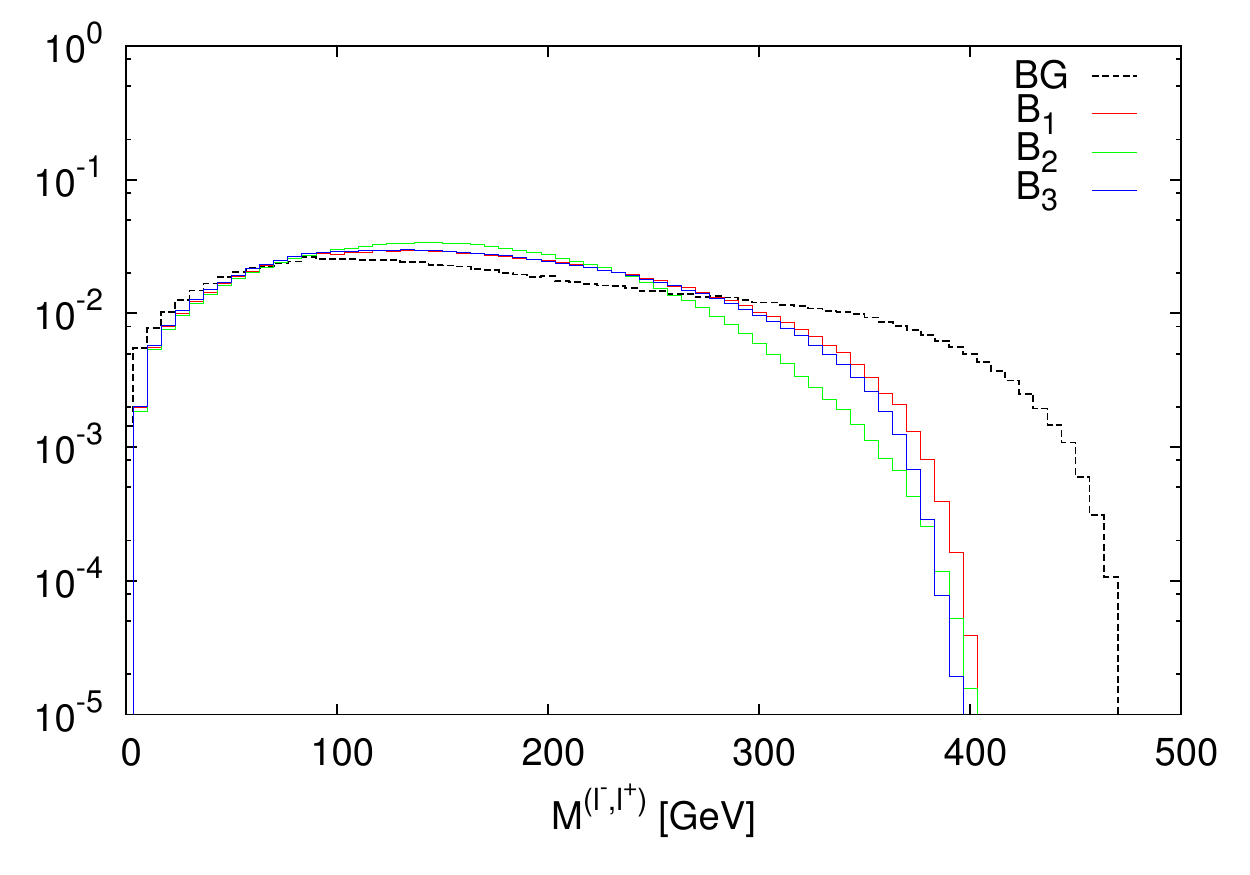}~\includegraphics[width=0.33\textwidth,height=0.18\textheight]{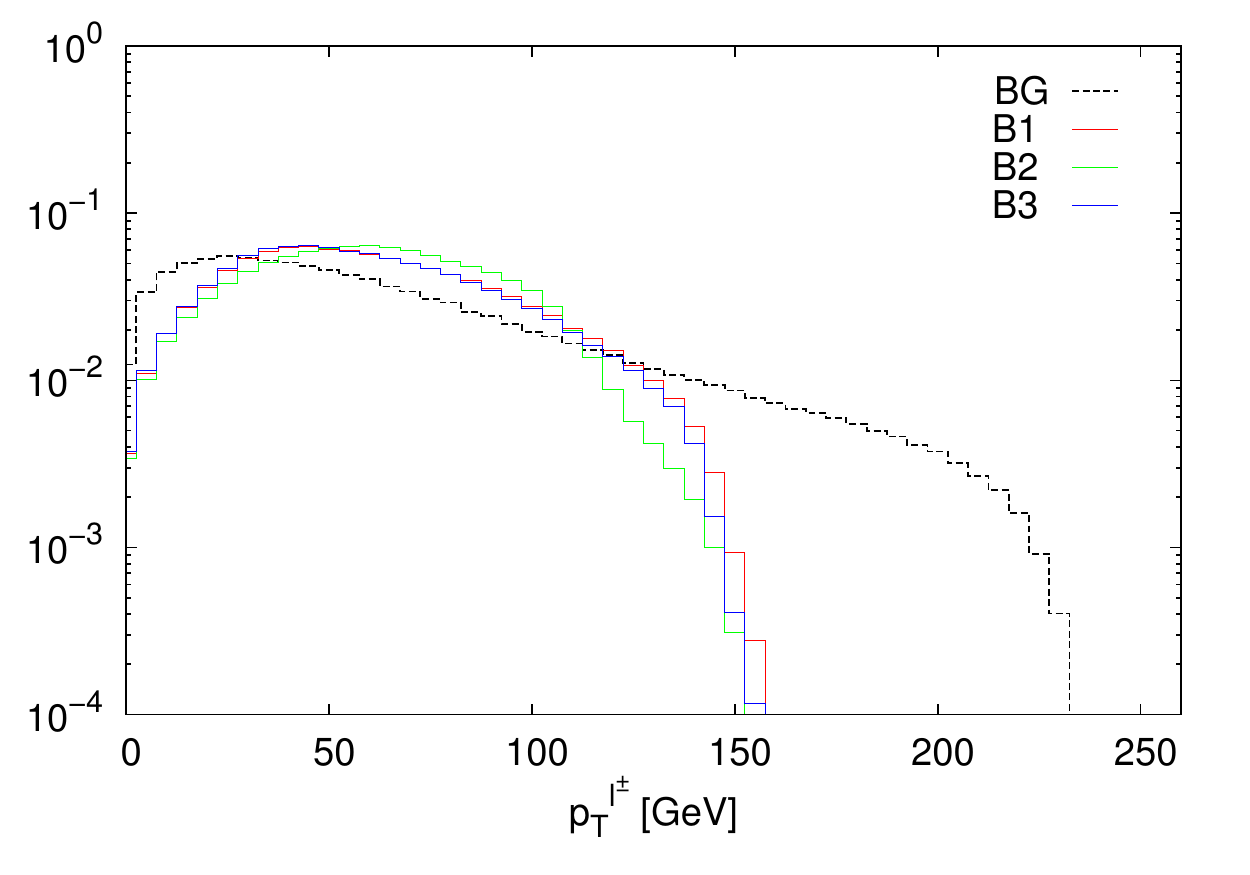}\\
\includegraphics[width=0.33\textwidth,height=0.18\textheight]{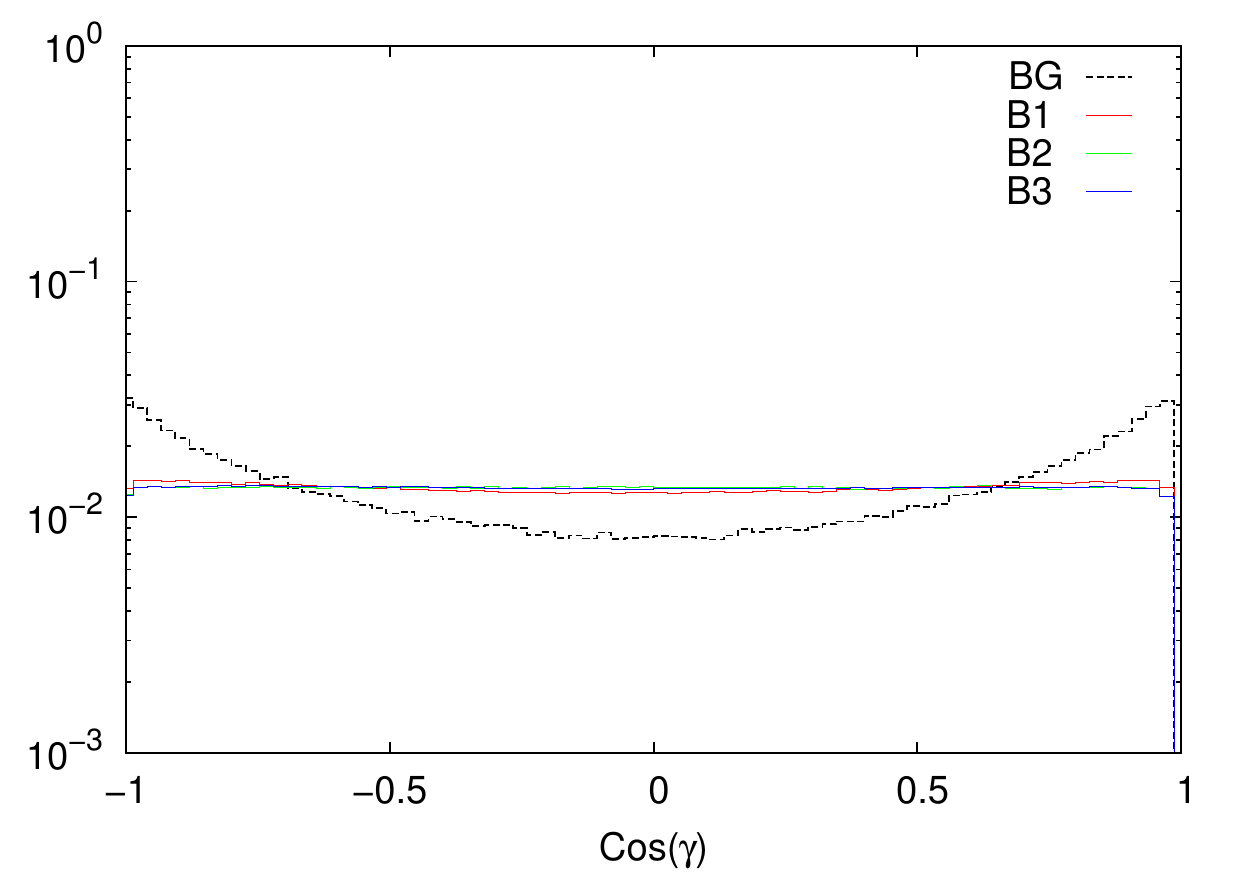}~\includegraphics[width=0.33\textwidth,height=0.18\textheight]{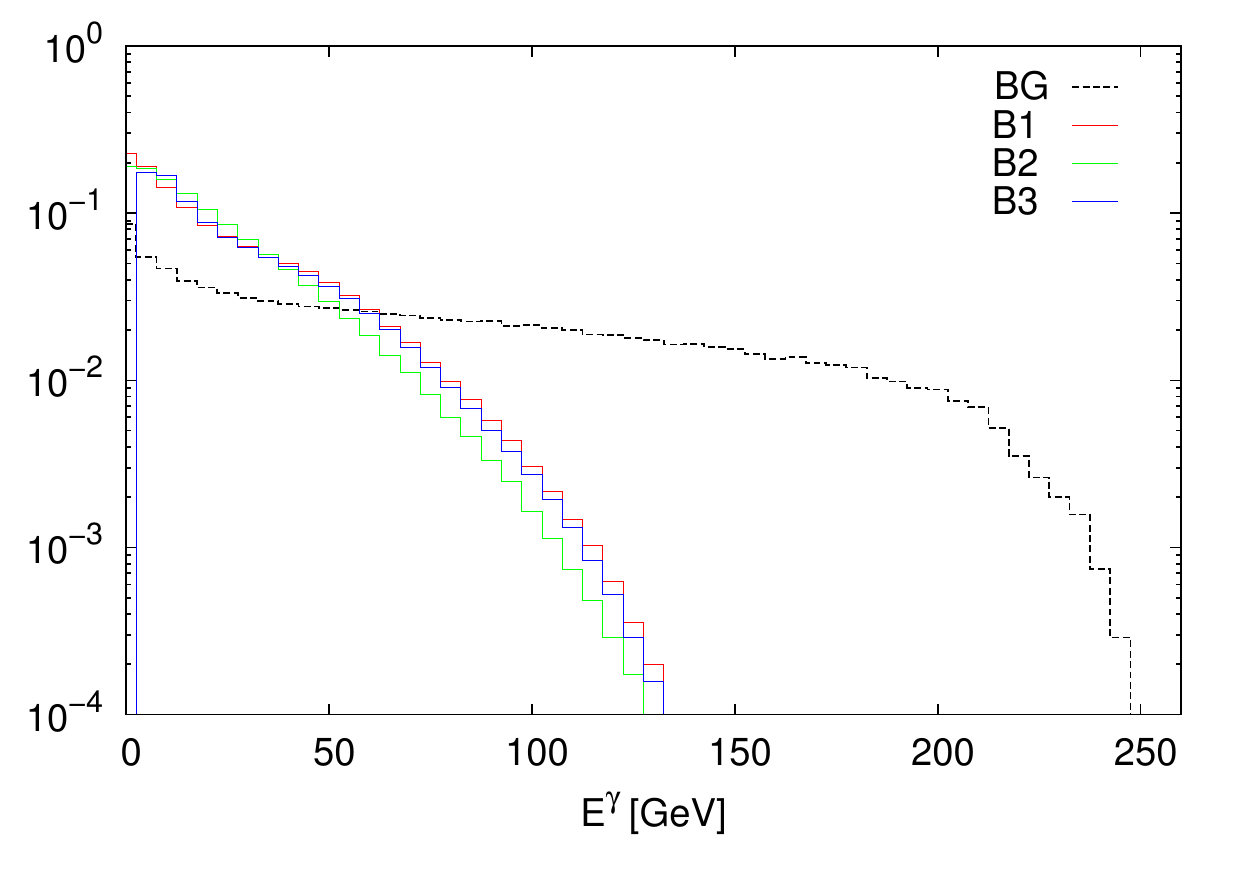}~\includegraphics[width=0.33\textwidth,height=0.18\textheight]{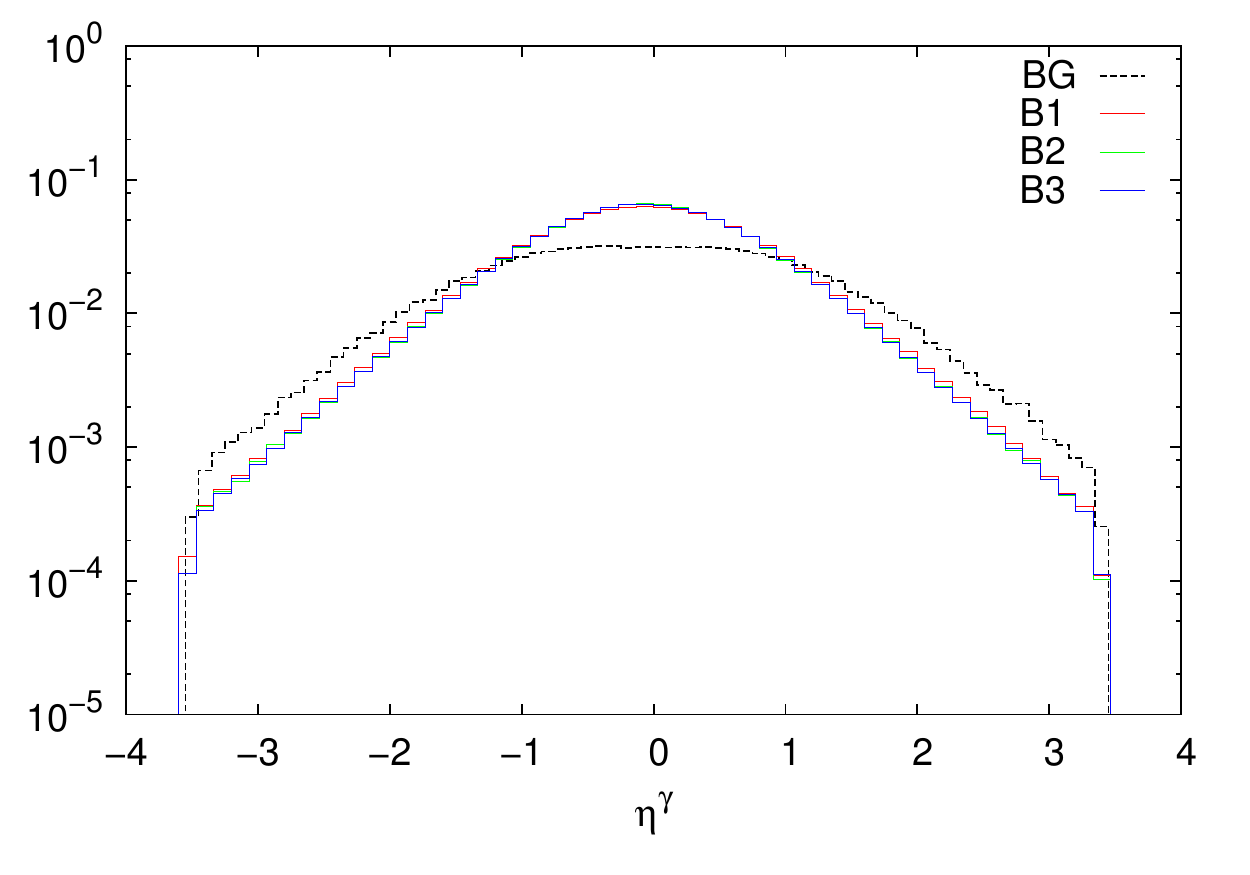} 
\par\end{centering}
\caption{Different normalized distributions of the process $e^{-}e^{+}\rightarrow S^{-}S^{+}+\gamma$ 
at $\sqrt{s}=500~\text{GeV}$. At the top from left to right, the angular normalized distribution 
$\textbf{cos}\left(\ell^{+},\ell^{-}\right)$, invariant mass $M_{\ell^{+},\ell^{-}}$, and the charged lepton transverse momentum 
$p_{t}^{\ell^{-}}$. At the bottom from left to right, the angular, the energy, and the rapidity normalized distributions
of the photon, respectively.}
\label{distr-eESA} 
\end{figure}

Similar to the monophoton analysis, we generate different distributions for the relevant kinematic variables and then select the following cuts that maximize the signal-to-background ratio: 

\begin{eqnarray}
\text{Final state}~S^{+}S^{-}:~\left\{
 \begin{array}{ll}
 M_{\ell^{+},\ell^{-}}<300~\text{GeV},~150~\text{GeV}<E_{miss}<420~\text{GeV},\nonumber\\
 30~\text{GeV}<E^{\ell}<180~\text{GeV},~p_t^{\ell}<170~\text{GeV},\label{cut-SS}
 \end{array}
 \right.
\end{eqnarray}
and
\begin{eqnarray}
\text{Final state}~S^{+}S^{-} \gamma:~\left\{
 \begin{array}{ll}
M_{\ell^{+},\ell^{-}}<300~\text{GeV},~150~\text{GeV}<E_{miss}<400~\text{GeV},\nonumber\\
30~\text{GeV}<E^{\ell}<170~\text{GeV},~p_t^{\ell}<170~\text{GeV},\nonumber\\
|\text{cos}\left(\theta_{\gamma}\right)|<0.5,~8~\text{GeV}<E^{\gamma}<120~\text{GeV},~p_{t}^{\gamma}<110~\text{GeV}.\label{cut-SSA}
 \end{array}
 \right.
\end{eqnarray}
The relevant normalized kinematical distributions for the processes $e^{-}e^{+}\rightarrow S^{-}S^{+}$
and $e^{-}e^{+}\rightarrow S^{-}S^{+} \gamma$ with the above cuts applied, are presented in Figs.~\ref{distr-eES} and \ref{distr-eESA}, respectively.

\begin{figure}[htp]
\begin{centering}
\includegraphics[width=0.48\textwidth,height=0.225\textheight]{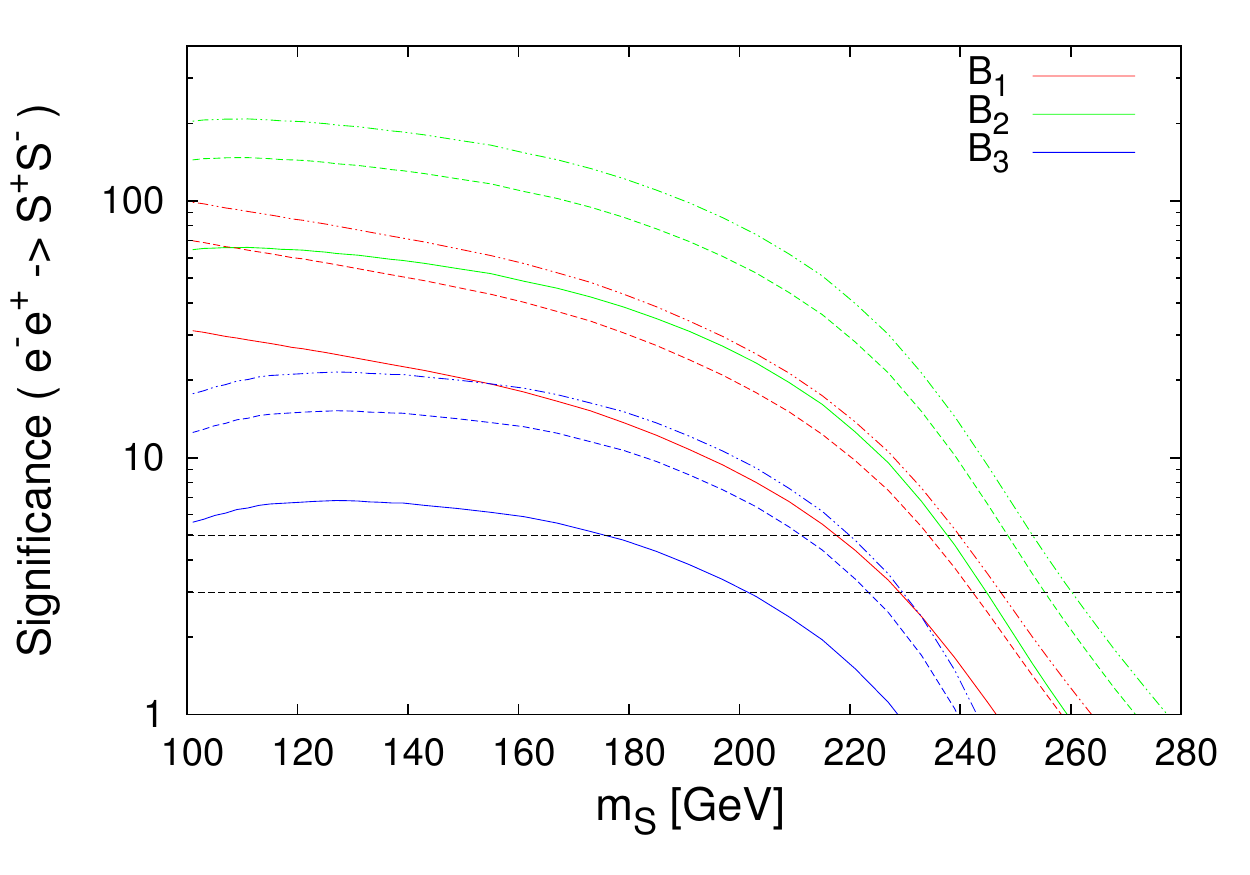}~\includegraphics[width=0.48\textwidth,height=0.225\textheight]{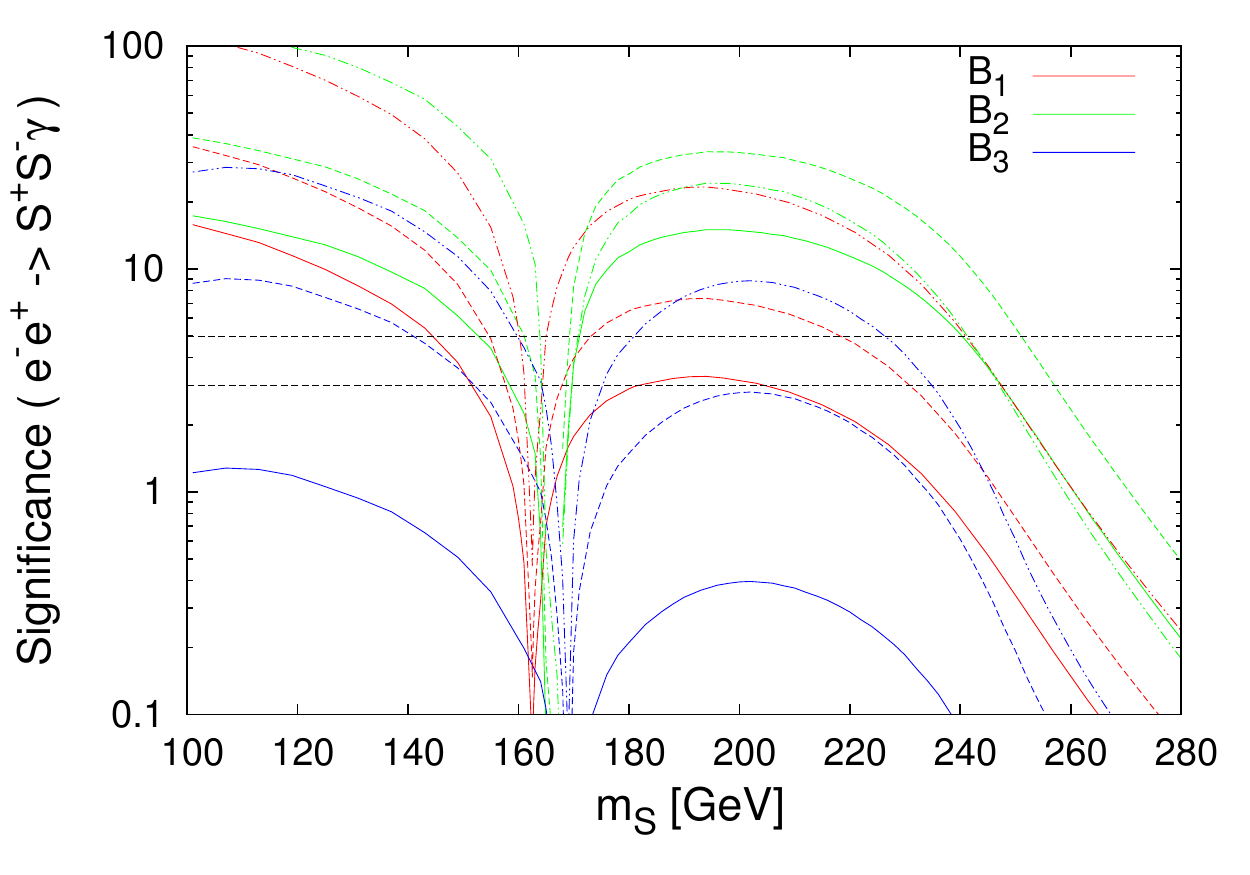} 
\par\end{centering}
\caption{The signal significance for the processes $e^{-}e^{+}\rightarrow S^{-}S^{+}$ (left) and $e^{-}e^{+}\rightarrow S^{-}S^{+}+\gamma$ (right) as a function of
$m_{S}$ for the values of $g_{i\alpha}$ given in Table-~\ref{tab-point} at an integrated luminosity of 1 (solid lines), 5 (dashed lines), and 10 $\text{fb}^{-1}$
 (dash-dotted lines) for $S^{-}S^{+}$ and an integrated luminosity of 10 (solid lines), 50 (dashed lines) and 100 $\text{fb}^{-1}$ (dash-dotted lines) 
for $S^{-}S^{+}+\gamma$. The horizontal dashed lines correspond to a 3 and 5 sigma significance. For the values $m_S> 250~\text{GeV}$, 
the charged scalars is off shell.}
\label{s-eES-A} 
\end{figure}

In Fig.~\ref{s-eES-A}, we show the significance for these processes as a function of $m_S$ for the three considered benchmark points. 
We see that for the production of a pair of charged scalars without a photon in the final state the signal-to-background ratio can be very 
large for $m_S < 220~\text{GeV}$ even at a very low integrated luminosity (about 0.5 $\text{fb}^{-1}$), whereas for larger masses the signal detection requires a huge luminosity. Hence, for charged scalars lighter than about $220~\text{GeV}$, the process $e^{-}e^{+}\rightarrow S^{-}S^{+}$ can be easily testable at the ILC. Concerning the final state with a photon, for a luminosity of few tens of $\text{fb}^{-1}$, the significance is large for a charged scalar mass less than $150~\text{GeV}$ and between $180$ and $220~\text{GeV}$.

\subsection{Analysis with polarized beams}

Using polarized electron or positron beams, which will be available at linear colliders such as the ILC, is an 
additional feature that allows the improvement of the detection of the signal of the considered processes. Indeed, 
for the processes we are considering, the interactions involve electrons (positrons) with only right- (left-) 
handed chirality, and a polarized beam can lead to an increase in the signal-to-background ratio. The ILC plans a 
longitudinal polarization of $80\%\left(|P\left(e^{-}\right)|<0.8\right)$ for the electron beam and 
$30\%\left(|P\left(e^{+}\right)|<0.3\right)$ for the positron beam with the possibility to upgrade to 
$60\%$. Here, we reanalyze the processes discussed earlier by considering polarized beams as 
$P\left(e^{-},e^{+}\right)=\left[+0.8,-0.3\right]$ while applying the same cuts used previously.

In Tables~\ref{tab-polariz-eEN}-\ref{tab-polariz-eESA}, we present the number of 
events of the signal and the background for the three benchmarks using different values of the electron and 
positron polarizations. We see clearly that with polarized beams the number of background events gets reduced 
by $86\%$ and the signal increased by $130\%$, and thus a substantial improvement of the significance for every 
process does exist. In particular, the process $e^{+}e^{-}\rightarrow S^{+}S^{-}+\gamma$ can be observed at a 
low luminosity of order of $\text{fb}^{-1}$ for partially polarized electron and positron beams, whereas it 
requires a large luminosity value without polarized beams.

\begin{table}[htp]
\begin{centering}
\begin{adjustbox}{max width=\textwidth} %
\begin{tabular}{|c|c|c|c|c|c|c|}
\hline 
$P(e^{-},e^{+})$ & $N_{BG}$ & BP & $N_{S}$ & $S_{10}$ & $S_{100}$ & $S_{500}$ \tabularnewline
\hline 
\hline 
\multirow{3}{*}{{[}0,0{]}} & \multirow{3}{*}{46652} & $B_{1}$ & 172.2 & 0.80 & 2.516 & 5.63 \tabularnewline
\cline{3-7} 
 & & $B_{2}$ & 122.2 & 0.565 & 1.79 & 3.99 \tabularnewline
\cline{3-7} 
 & & $B_{3}$ & 130.1 & 0.61 & 1.90 & 4.25 \tabularnewline
\hline 
\multirow{3}{*}{{[}+0.8,-0.3{]}} & \multirow{3}{*}{6541} & $B_{1}$ & 396.06 & 4.75 & 15.04 & 33.62 \tabularnewline
\cline{3-7} 
 & & $B_{2}$ & 283.5 & 3.43 & 10.85 & 24.27 \tabularnewline
\cline{3-7} 
 & & $B_{3}$ & 299.23 & 3.62 & 11.44 & 25.58 \tabularnewline
\hline 
\end{tabular}\end{adjustbox} 
\par\end{centering}
\caption{The background ($N_{BG}$) and signal ($N_{S}$) number of events
for the process $e^{-}e^{+}\rightarrow\gamma+E_{miss}$ that corresponds
to the integrated luminosity values L=10 $\text{fb}^{-1}$ for the three chosen
benchmark points (Table-\ref{tab-point}), with or without polarized
beams, within the cuts given in (\ref{cut-monoA}); the significance
values $S_{10}$, $S_{100}$ and $S_{500}$ correspond to the
three integrated luminosity values $L=10,~100,~\text{and}~500~\text{fb}^{-1}$.}
\label{tab-polariz-eEN} 
\end{table}

\begin{table}[htp]
\begin{centering}
\begin{adjustbox}{max width=\textwidth} %
\begin{tabular}{|c|c|c|c|c|c|c|}
\hline 
$P(e^{-},e^{+})$ & $N_{BG}$ & BP & $N_{S}$ & $S_{0.1}$ & $S_{0.5}$ & $S_{1}$ \tabularnewline
\hline 
\hline 
\multirow{3}{*}{{[}0,0{]}} & \multirow{3}{*}{212036} & $B_{1}$ & 11312 & 2.39 & 5.35 & 7.57 \tabularnewline
\cline{3-7} 
 & & $B_{2}$ & 7231 & 1.54 & 3.45 & 4.88 \tabularnewline
\cline{3-7} 
 & & $B_{3}$ & 10660 & 2.26 & 5.05 & 7.14 \tabularnewline
\hline 
\multirow{3}{*}{{[}+0.8,-0.3{]}} & \multirow{3}{*}{122397} & $B_{1}$ & 25904 & 6.73 & 15.04 & 21.27 \tabularnewline
\cline{3-7} 
 & & $B_{2}$ & 17138 & 4.59 & 10.26 & 14.51 \tabularnewline
\cline{3-7} 
 & & $B_{3}$ & 24625 & 6.42 & 14.36 & 20.31 \tabularnewline
\hline 
\end{tabular}\end{adjustbox} 
\par\end{centering}
\caption{The background ($N_{BG}$) and signal ($N_{S}$) number of events
for the process $e^{-}e^{+}\rightarrow S^{-}S^{+}$ that corresponds
to the integrated luminosity values L=10 $\text{fb}^{-1}$ for the three chosen
benchmark points (Table-\ref{tab-point}), with or without polarized
beams, within the cuts given in (\ref{cut-SS}); the significance
values $S_{0.1}$, $S_{0.5}$, and $S_{1}$ correspond to the
three integrated luminosity values $L=0.1,~0.5,~\text{and}~1~\text{fb}^{-1}$.}
\label{tab-polariz-eES} 
\end{table}

\begin{table}[htp]
\begin{centering}
\begin{adjustbox}{max width=\textwidth} %
\begin{tabular}{|c|c|c|c|c|c|c|}
\hline 
$P(e^{-},e^{+})$ & $N_{BG}$ & BP & $N_{S}$ & $S_{10}$ & $S_{50}$ & $S_{100}$ \tabularnewline
\hline 
\hline 
\multirow{3}{*}{{[}0,0{]}} & \multirow{3}{*}{876.39} & $B_{1}$ & 26.56 & 0.88 & 1.98 & 2.79 \tabularnewline
\cline{3-7} 
 & & $B_{2}$ & 10.52 & 0.35 & 0.79 & 1.12 \tabularnewline
\cline{3-7} 
 & & $B_{3}$ & 66.10 & 2.15 & 4.81 & 6.81 \tabularnewline
\hline 
\multirow{3}{*}{{[}+0.8,-0.3{]}} & \multirow{3}{*}{123.20} & $B_{1}$ & 61.48 & 4.52 & 10.11 & 14.30 \tabularnewline
\cline{3-7} 
 & & $B_{2}$ & 24.24 & 2.00 & 4.46 & 6.31 \tabularnewline
\cline{3-7} 
 & & $B_{3}$ & 150.05 & 9.08 & 20.30 & 28.70 \tabularnewline
\hline 
\end{tabular}\end{adjustbox} 
\par\end{centering}
\caption{The background ($N_{BG}$) and signal ($N_{S}$) number of events
for the process $e^{-}e^{+}\rightarrow S^{-}S^{+}+\gamma$ that corresponds
to the integrated luminosity values L=10 $\text{fb}^{-1}$ for the three chosen
benchmark points (Table-\ref{tab-point}), with or without polarized
beams, within the cuts given in (\ref{cut-SSA}); the significance
values $S_{10}$, $S_{50}$ and $S_{100}$ correspond to the three
integrated luminosity values $L=10,~50,~\text{and}~100~\text{fb}^{-1}$.}
\label{tab-polariz-eESA} 
\end{table}

\begin{figure}[htp]
\begin{centering}
\includegraphics[width=0.48\textwidth,height=0.26\textheight]{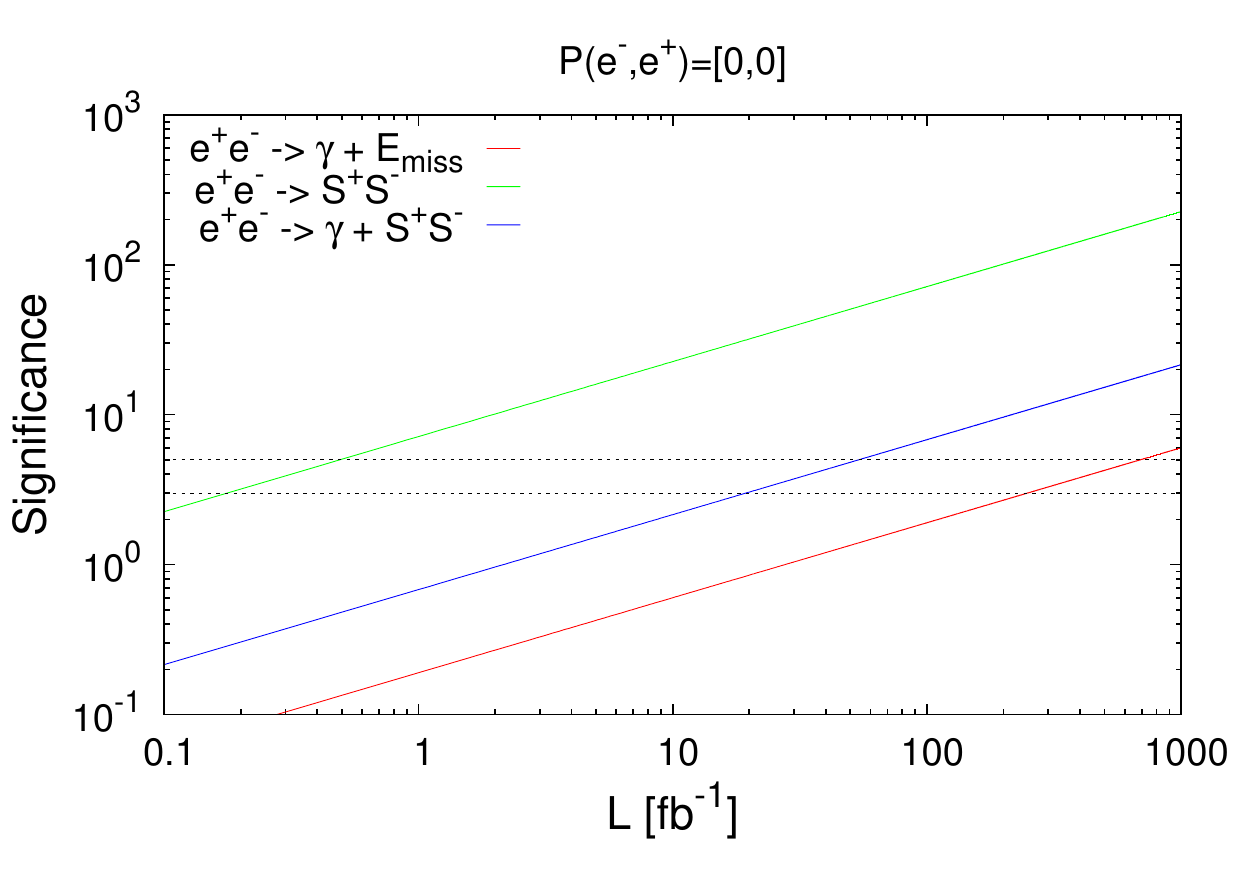}~\includegraphics[width=0.48\textwidth,height=0.26\textheight]{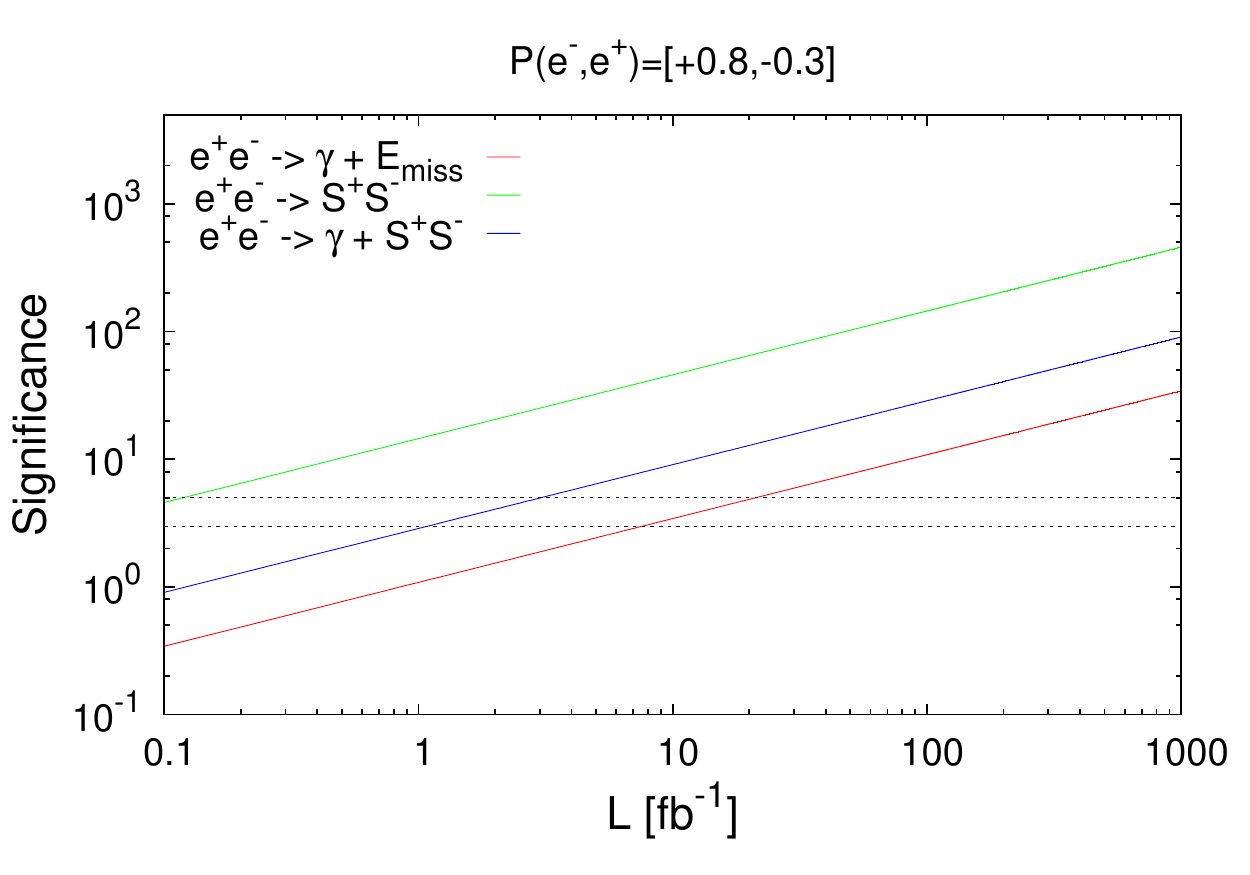} 
\par\end{centering}
\caption{The signal significance versus the luminosity for the three different signatures studied, without polarization at left and with polarization of $P(e^{-},e^{+})$=$[+0.8,-0.3]$ at right. The two horizontal dashed lines in each figure indicate the corresponding significance values for $S = 3$ and $S = 5$, respectively.}
\label{sig-lum} 
\end{figure}

In order to see the effect of the polarization on the signal, we present in Fig.~\ref{sig-lum} 
the significance for $P(e^{-},e^{+})$=$[0,0]$ and $P(e^{-},e^{+})$=$[+0.8,-0.3]$, for the benchmark point $B_3$ as a function of luminosity. 
We clearly see that the signal over the background gets improved significantly. For $P(e^{-},e^{+})=[-0.8,+0.3]$, 
$[N_{BG},~N_S]$ %the events numbers of signal and background 
get modified by $[+189\%,-88\%]$. In the case where only the electron beam is polarized with $P(e^{-})$=$+0.8 (-0.8)$, $[N_{BG},~N_S]$ 
are changed by $[-78\%,+80\%]$ ($[+85\%,-78\%])$. On the other hand, where only the positron beam is polarized with 
$P(e^{+})$=$+0.3(-0.3)$, $[N_{BG},~N_S]$ are changed by $[+63\%,-43\%]$ ($[-59\%,+62\%]$).

\section{Conclusion}

In this work, we investigated some of type of interactions that are part of a generic class of radiative 
neutrino mass models which involve a heavy right-handed neutrino coupled to a RH charged lepton and singlet 
charged scalar. We find that in order to be consistent with the current experimental bounds on the LFV 
processes such as $\ell_{\alpha}\rightarrow \ell_{\beta}+\gamma$ requires the coupling of the charged 
leptons with the RH neutrinos to be suppressed which can be in conflict with the observed DM relic density. 
For that we defined a fine-tuning parameter $R$ that measures how small the couplings have to be to satisfy 
both the DM constraint and the LFV bounds. Hence, in our analysis we consider three sets of benchmark points that 
avoid the experimental limits on the LFV branching fraction with $R_1\sim~1$ (without fine-tuning), 
$R_2\sim~10^{-2}$ (moderately fine-tuned), and $R_2\sim~10^{-4}$ (highly fine-tuned). We also used the data 
from LEP-II on a monophoton plus missing energy to constrain the model space parameter and derived an 
approximate analytical bound on the coupling of the right-handed charged lepton to the RH neutrino and 
charged scalar. Interestingly, we found that, even before applying kinematical cuts, there are points for 
which the cross sections of the processes $e^-e^+\rightarrow \gamma+E_{miss},~S^-S^+,~S^-S^++\gamma$ are 
much larger than the corresponding background by more than one order of magnitude. Thus, the future lepton 
colliders will be able to probe a significant fraction of the parameter space of this class of radiative neutrino 
mass models even at a low luminosity.

Among the scanned benchmark points with comparable RH neutrino masses, we have chosen three of them with 
$ R_1 \approx 1, R_2 \approx 10^{-2}$, and $ R_3 \approx 10^{-4}$ to see the effect of the fine-tuning. Then 
we applied the appropriate cuts on the kinematical variables to reduce the background. We found that 
the signal can be seen at an integrated luminosity of $\mathcal{O}(100~\text{fb}^{-1}),~\mathcal{O}(\text{fb}^{-1}), \text{and}~\mathcal{O}(10~\text{fb}^{-1})$ 
for the processes $\gamma + E_{miss}$, $S^+S^-$, and $S^+S^-+\gamma$, respectively, whereas, one cannot disentangle 
the effect of the fine-tuning. For these values of the 
luminosity, the charged scalar should be lighter than $220~\text{GeV}$.

Finally, since the interactions considered in this work involve exclusively the right-handed charged lepton, we 
studied the effect of polarized $e^{-}/e^{+}$ beams, as will be used at future linear colliders such as the ILC 
or CLIC. We have shown that the signal-to-background ratio gets enhanced by a factor of more than 5 for the polarization
 $P(e^{-},e^{+})=[+0.8,-0.3]$. Thus, with the use of polarized beams, a large parameter space of this class of neutrino 
radiative models can be probed for with a low luminosity via different processes at the starting of the planned ILC and CLIC colliders.

\appendix
\section{Loop functions \label{appsec:appendix}}

Here, we give the loop functions used in Sec. II, which are given
by

\begin{align}
F\left(x\right) & =\frac{1-6x+3x^{2}+2x^{3}-6x^{2}\log x}{6\left(1-x\right)^{4}},\\
G\left(x\right)= & \frac{2-9x+18x^{2}-11x^{3}+6x^{3}\log x}{6\left(1-x\right)^{4}},\\
D_{1}\left(x,y\right)= & -\frac{1}{\left(1-x\right)\left(1-y\right)}-\frac{x^{2}\log x}{\left(1-x\right)^{2}\left(x-y\right)}-\frac{y^{2}\log y}{\left(1-y\right)^{2}\left(y-x\right)},\\
D_{2}\left(x,y\right)= & -\frac{1}{\left(1-x\right)\left(1-y\right)}-\frac{x\log x}{\left(1-x\right)^{2}\left(x-y\right)}-\frac{y\log y}{\left(1-y\right)^{2}\left(y-x\right)}.
\end{align}
These loop functions does not diverge and behave as follows near critical
points:
\begin{align}
D_{1}\left(x,x\right) & =\frac{-1+x^{2}-2x\log x}{\left(1-x\right)^{3}},\\
D_{2}\left(x,x\right)= & \frac{-2+2x-\left(1+x\right)\log x}{\left(1-x\right)^{3}},\\
D_{1}\left(x,1\right)=D_{1}\left(1,x\right)= & \frac{-1+4x-3x^{2}+2x^{2}\log x}{2\left(1-x\right)^{3}},\\
D_{2}\left(x,1\right)=D_{2}\left(1,x\right) & =\frac{1-x^{2}+2x\log x}{2\left(1-x\right)^{3}},
\end{align}
and
\begin{equation}
F\left(1\right)=\frac{1}{10},\qquad G\left(1\right)=\frac{1}{4},\qquad D_{1}\left(1,1\right)=\frac{1}{3},\qquad D_{2}\left(1,1\right)=\frac{1}{6}.
\end{equation}

\end{document}